\input harvmac
\input amssym.def
\input amssym.tex

\parskip=4pt \baselineskip=12pt
\hfuzz=20pt
\parindent 10pt

\def\ta{{\tilde\alpha}}
\def\td{{\tilde d}}
\def\te{{\tilde e}}
\def\tf{{\tilde f}}
\def\tg{{\tilde g}}
\def\th{{\tilde h}}
\def\tk{{\tilde k}}
\def\hh{{\hat h}}
\def\hk{{\hat k}}

\def\nt{\noindent}
\def\nl{\hfill\break}

\def\np{\vfill\eject}

\def\hel{{\textstyle{11\over2}}}

\global\newcount\subsubsecno \global\subsubsecno=0
\def\subsubsec#1{\global\advance\subsubsecno
by1\message{(\secsym\the\subsecno.\the\subsubsecno. #1)}
\ifnum\lastpenalty>9000\else\bigbreak\fi
\noindent{\bf\secsym\the\subsecno.\the\subsubsecno. #1}\writetoca{\string\quad
{\secsym\the\subsecno.\the\subsubsecno.} {#1}}\par\nobreak\medskip\nobreak}

\def\subsubsec#1{\global\advance\subsubsecno
by1\message{(\secsym\the\subsecno.\the\subsubsecno.
#1)} \ifnum\lastpenalty>9000\else\bigbreak\fi
\noindent{\bf\secsym\the\subsecno.\the\subsubsecno. #1}\writetoca{\string\quad
{\secsym\the\subsecno.\the\subsubsecno.} {#1}}}%

%%%%%%%%%%%%%%%%%%%%%%%%%%%%%%%%%%%%%%%%%%%%%%%%%%%%%%%%%%%%%%%%
\input epsf.tex
%%%%%%%%%%%%%%%%%%%%%% macros for figures %%%%%%%%%%%%%%%%%%%%%%%
\newcount\figno
\figno=0
\def\fig#1#2#3{
\par\begingroup\parindent=0pt\leftskip=1cm\rightskip=1cm\parindent=0pt
\baselineskip=11pt \global\advance\figno by 1 \midinsert
\epsfxsize=#3 \centerline{\epsfbox{#2}} \vskip 12pt
%{\bf Fig. \the\figno:}
#1\par
\endinsert\endgroup\par}
\def\figlabel#1{\xdef#1{\the\figno}}
\def\encadremath#1{\vbox{\hrule\hbox{\vrule\kern8pt\vbox{\kern8pt
\hbox{$\displaystyle #1$}\kern8pt} \kern8pt\vrule}\hrule}}
%%%%%%%%%%%%%%%%%%%%%%%%%%%%%%%%%%%%%%%%%%%%%%%%%%%%%%%%%%%%%%%%%

  \def\tV{{\tilde V}}

\def\rank{{\rm rank}}
\def\downcirc#1{\mathop{\circ}\limits_{#1}}
\def\riga{-\kern-4pt - \kern-4pt -}
\font\fat=cmsy10 scaled\magstep5

\def\Bbullet{\raise-3pt\hbox{\fat\char"0F}}

\def\black#1{\mathop{\bullet}\limits_{#1}}

\font\tfont=cmbx12 scaled\magstep1 %large
\font\male=cmr9

\def\Box{
\vbox{ \halign to5pt{\strut##& \hfil ## \hfil \cr &$\kern -0.5pt
\sqcap$ \cr \noalign{\kern -5pt \hrule} }}~}

\def\down{\raise1.5pt\hbox{$\phantom{a}_2$}\downarrow}

\def\downa{\raise1.5pt\hbox{$\phantom{a}_{2\atop m_2}$}\downarrow}

\def\llr{\longrightarrow}

\def\({\left(}
\def\){\right)}

\def\lra{\longrightarrow}
\def\llra{\longleftrightarrow}

\def\ha{{\textstyle{1\over2}}}

\def\bbc{{C\kern-6.5pt I}}
\def\bac{{C\kern-5.5pt I}}
\def\bab{{C\kern-4.5pt I}}

\def\bbr{{I\!\!R}}
\def\bbn{I\!\!N}
\def\a{\alpha}
\def\b{\beta}

\def\vr{\vert}

\def\l{\lambda}

\def\ca{{\cal A}}  \def\cc{{\cal C}}
\def\cd{{\cal D}} \def\ce{{\cal E}} \def\cf{{\cal F}}
\def\cg{{\cal G}} \def\ch{{\cal H}} 
 \def\ck{{\cal K}} 
\def\cm{{\cal M}} \def\cn{{\cal N}} 
\def\cp{{\cal P}}  
 \def\ct{{\cal T}}

\def\ido{intertwining differential operator}
\def\idos{intertwining differential operators}

\def\L{\Lambda}
\def\r{\rho}

%%%%%%%%%%%%%%

%%%%% \lrefs

\lref\KanLan{J. Kang and P. Langacker, Phys. Rev. {\bf D71} (2005) 035014,
hep-ph/0412190.}
%Z' Discovery Limits For Supersymmetric E_6 Models

\lref\GanVas{T. Gannon and M. Vasudevan, JHEP 0507:035 (2005),
hep-th/0504006.}
%Charges of Exceptionally Twisted Branes

\lref\Ohwa{Y. Ohwashi, Prog. Theor. Phys. {\bf 115}, 625-659 (2006),
 hep-th/0510252.}
%Sp(4,H)/Z_2 Pair Universe in E6 Matrix Models

\lref\HunMos{P.Q. Hung and Paola Mosconi, hep-ph/0611001.}
%$E_6$ unification in a model of dark energy and dark matter

\lref\DufFer{M.J. Duff and S. Ferrara, Phys. Rev. {\bf D76}, 124023 (2007),
arXiv:0704.0507v2 [hep-th].}
%E_6 and the bipartite entanglement of three qutrits

\lref\FaKoRi{A.E. Faraggi, C. Kounnas and J. Rizos,
Nucl. Phys. {\bf B774}, 208-231 (2007), hep-th/0611251.}
%Spinor-Vector Duality in fermionic Z2XZ2 heterotic orbifold models

\lref\GuLuMi{S. Gurrieri, A. Lukas and A. Micu,
JHEP 0712:081 (2007), arXiv:0709.1932.}
% Heterotic String Compactifications on Half-flat Manifolds II

\lref\BCCS{F. Bernardoni, S.L. Cacciatori, Bianca L. Cerchiai and A. Scotti,
J. Math. Phys. {\bf 49}, 012107 (2008), arXiv:0710.0356.}
%Mapping the geometry of the E6 group

\lref\Kallosh{R. Kallosh and M. Soroush,
%Explicit Action of E7(7) on N=8 Supergravity Fields
Nucl. Phys. {\bf B801}, 25-44 (2008), arXiv:0802.4106;\nl
R. Kallosh and T. Kugo, arXiv:0811.3414.}

\lref\Mizo{Sh. Mizoguchi, arXiv:0808.2857.}
% Localized Modes in Type II and Heterotic Singular Calabi-Yau Conformal Field Theories

\lref\HowKin{R. Howl and S.F. King, JHEP, 0801:060 (2008).}
%Minimal E_6 Supersymmetric Standard Model. .

\lref\Helg{S. Helgason, {\it Differential Geometry, Lie Groups
and Symmetric Spaces},\hfill\break (Academic Press, New York,
1978).}

\lref\Dobmul{V.K. Dobrev, Lett. Math. Phys. {\bf 9}, 205-211 (1985).}
%Multiplet classification of the reducible elementary
%representations of real semi-{\allowbreak}simple Lie groups: the \
%$SO_e(p,q)$ example,

\lref\Gilm{R. Gilmore, {\it Lie groups, Lie algebras, and some of
their applications}, (New York, Wiley, 1974).}

\lref\BaRo{A.O. Barut and R. R\c aczka, {\it
Theory of Group Representations and Applications}, \hfil\break II
edition, (Polish Sci. Publ., Warsaw, 1980).}

\lref\Ter{J. Terning, {\it Modern Supersymmetry: Dynamics and
Duality}, International Series of Monographs on Physics \# 132,
(Oxford University Press, 2005).}

\lref\Har{Harish-Chandra, ``Discrete series for semisimple Lie groups:
II'', Ann. Math. {\bf 116} (1966) 1-111.}

\lref\KnSt{A.W. Knapp and E.M. Stein, ``Intertwining operators for
semisimple groups'', Ann. Math. {\bf 93} (1971) 489-578; II : Inv.
Math. {\bf 60} (1980) 9-84.}

\lref\BGG{I.N. Bernstein, I.M. Gel'fand and S.I. Gel'fand,
``Structure of representations generated by highest weight vectors'',
Funkts. Anal. Prilozh. {\bf 5} (1) (1971) 1-9; English translation:
Funct. Anal. Appl. {\bf 5} (1971) 1-8.}

\lref\War{G. Warner, {\it Harmonic Analysis on Semi-Simple Lie
Groups I}, (Springer, Berlin, 1972).}

\lref\Lan{R.P. Langlands, {\it On the classification of irreducible
representations of real algebraic groups}, Math. Surveys and
Monographs, Vol. 31 (AMS, 1988), first as IAS Princeton preprint
(1973).}

\lref\Zhea{D.P. Zhelobenko, {\it Harmonic Analysis on Semisimple
Complex Lie Groups}, (Moscow, Nauka, 1974, in Russian).}

\lref\Dix{J. Dixmier, {\it Enveloping Algebras}, (North Holland, New
York, 1977).}

\lref\Bourb{N. Bourbaki, {\it Groupes at alg\`{e}bres de Lie,
Chapitres 4,5 et 6}, (Hermann, Paris, 1968).}

\lref\KnZu{A.W. Knapp and G.J. Zuckerman,
``Classification theorems for representations of semisimple groups'',
 in: Lecture Notes in Math., Vol. 587 (Springer, Berlin,
1977) pp. 138-159; ~``Classification of irreducible tempered representations of semisimple groups'',
Ann. Math. {\bf 116} (1982) 389-501.}

\lref\DMPPT{V.K. Dobrev, G. Mack, V.B. Petkova, S.G. Petrova and
I.T. Todorov, {\it Harmonic Analysis on the $n$-Dimensional Lorentz
Group and Its Applications to Conformal Quantum Field Theory},
Lecture Notes in Physics, Vol. 63 (Springer-Verlag,
 Berlin-Heidelberg-New York, 1977);
V.K. Dobrev and V.B. Petkova, Rept. Math. Phys. {\bf 13}, 233-277 (1978).}

\lref\Knapp{A.W. Knapp, {\it Representation Theory of Semisimple
Groups (An Overview Based on Examples)}, (Princeton Univ. Press,
1986).}

\lref\Dob{V.K. Dobrev,
%``Canonical construction of intertwining differential operators
%associated with representations of real semisimple Lie groups'',
Rept. Math. Phys. {\bf 25}, 159-181 (1988) ; first as ICTP Trieste
preprint IC/86/393 (1986).}

\lref\DobPet{V.K. Dobrev and V.B. Petkova,
Phys. Lett. {\bf B162}, 127-132 (1985);
Lett. Math. Phys. {\bf 9}, 287-298 (1985);
Fortsch. Phys. {\bf 35}, 537-572 (1987).}

\lref\Dobsusy{V.K. Dobrev,
Phys. Lett. {\bf B186}, 43-51 (1987);
% and ICTP Trieste preprint IC/86/241
J. Phys. {\bf A35}, 7079-7100 (2002),
hep-th/0201076; Phys. Part. Nucl. {\bf 38}, 564-609 (2007),
 hep-th/0406154;
V.K. Dobrev and A.Ch. Ganchev,
Mod. Phys. Lett. {\bf A3}, 127-137 (1988).}

\lref\Dobso{V.K. Dobrev,
%``Invariant differential operators and characters of the $AdS_4$ algebra'',
J. Phys. {\bf A39} (2006) 5995-6020; hep-th/0512354.}

\lref\Dobsin{V.K. Dobrev,
%``Singular vectors of quantum groups
%representations for straight Lie algebra roots'',
Lett. Math. Phys. {\bf 22}, 251-266 (1991).}

\lref\Dobqg{V.K. Dobrev,
J. Math. Phys. {\bf 33}, 3419-3430 (1992) %IC-92-18, Jan 1992;
J. Phys. {\bf A26}, 1317-1334 (1993), first as G\"{o}ttingen
University preprint, (July 1991);
~%``$q$ - difference intertwining operators for
%$U_q(sl(n))$: general setting and the case $n=3$'',
J. Phys. {\bf A27}, 4841-4857 (1994), Erratum-ibid. {\bf A27},
6633-6634 (1994), hep-th/9405150;
~ % ``New $q$ - Minkowski space-time and $q$ - Maxwell
%equations hierarchy from $q$ - conformal invariance'',
Phys. Lett. {\bf B341}, 133-138 (1994) Erratum-ibid. {\bf B346}, 427 (1995);
~V.K. Dobrev and P.J. Moylan, %``Finite-dimensional singletons of
%the quantum anti de Sitter algebra'',
Phys. Lett. {\bf B315}, 292-298 (1993) .} %ICTP Trieste IC/93/059 (March 1993).}

\lref\Sata{I. Satake,
%``On representations and compactifications of symmetric Riemannian spaces'',
Ann. Math. {\bf 71} (1960) 77-110.}

%\lref\Dobp{V.K. Dobrev, in preparation.}

\lref\Dobinv{V.K. Dobrev, %Invariant Differential Operators for Non-Compact
%Lie Groups: Parabolic Subalgebras,
Rev. Math. Phys. {\bf 20} (2008)
407-449; hep-th/0702152; ICTP Trieste preprint IC/2007/015.}

\lref\Dobpeds{V.K. Dobrev, %Positive Energy Representations, Holomorphic
%Discrete Series and Finite-Dimensional Irreps,
J. Phys. A: Math. Theor. {\bf 41} (2008) 425206; arXiv:0712.4375 [hep-th].}

%%%%

\lref\EHW{T. Enright, R. Howe and W. Wallach, "A classification of
unitary highest weight modules", in: {\it Representations of
Reductive Groups}, ed. P. Trombi (Birkh\"auser, Boston, 1983) pp.
97-143.}

\lref\Witten{E. Witten, "Conformal Field Theory in Four and Six
Dimensions", arXiv:0712.0157.}

%% START OF TEXT

%\rightline{INRNE-TH-08-12}

\vskip 1.5cm

\centerline{{\tfont Invariant Differential Operators}} \vskip
2truemm \centerline{{\tfont for Non-Compact Lie Groups:}} \vskip
2truemm \centerline{{\tfont the E$_{\hbox{\bf 6(-14)}}$ case}
\foot{Invited talks at V School in Modern Mathematical Physics, Belgrade, Serbia,
6-17.7.2008.}}

\vskip 1.5cm

\centerline{{\bf V.K. Dobrev}}
\vskip 0.5cm

\centerline{Institute for Nuclear Research and Nuclear Energy}
\centerline{Bulgarian Academy of Sciences} \centerline{72
Tsarigradsko Chaussee, 1784 Sofia, Bulgaria} \centerline{(permanent
address)}
%dobrev@inrne.bas.bg

\vskip 0.5cm \centerline{and}

\vskip 0.5cm

 \centerline{The Abdus Salam International Centre for
Theoretical Physics} \centerline{P.O. Box 586, Strada Costiera 11}
\centerline{34014 Trieste, Italy}

\vskip 1.5cm

 \centerline{{\bf Abstract}}
\midinsert\narrower{\male In the present paper we continue the project of
systematic explicit construction of invariant differential operators.
On the example of the non-compact exceptional group $E_{6(-14)}$
we give the multiplets of indecomposable elementary representations.
This includes the data for all relevant invariant differential operators.
 }\endinsert

\vskip 1.5cm

\newsec{Introduction}

\nt Invariant differential operators play very important role in the
description of physical symmetries - starting from the early
occurrences in the Maxwell, d'Allembert, Dirac, equations, (for more
examples cf., e.g., \BaRo{}), to the latest applications of
(super-)differential operators in conformal field theory,
supergravity and string theory, (for a recent review, cf. e.g.,
\Ter). Thus, it is important for the applications in physics to
study systematically such operators.

In a recent paper \Dobinv{} we started the systematic explicit
construction of invariant differential operators. We gave an
explicit description of the building blocks, namely, the parabolic
subgroups and subalgebras from which the necessary representations
are induced. Thus we have set the stage for study of different
non-compact groups.

In the present paper we focus on one particular group
~$E_{6(-14)}\,$, which is very interesting for at least two reasons.
First of all exceptional groups are still not much studied and used,
cf. though
\KanLan,\GanVas,\Ohwa,\HunMos,\DufFer,\FaKoRi,\GuLuMi,\BCCS,\Kallosh,\Mizo,\HowKin.
Furthermore, ~$E_{6(-14)}\,$~ is one of two exceptional non-compact
groups that have highest/lowest weight representations.\foot{The
other one is ~$E_{7(-25)}\,$ which we also plan to consider.}

In our further plans it shall be very useful that (as in \Dobinv) we
 follow a procedure in representation
theory in which \idos\ appear canonically \Dob{} and which procedure
has been generalized to the supersymmetry setting \DobPet,\Dobsusy{}
and to quantum groups \Dobqg. (For more references, cf. \Dobinv.)

The present paper is organized a follows. In section 2 we give the
preliminaries, actually recalling and adapting facts from \Dobinv.
In Section 3 we specialize to the ~$E_{6(-14)}\,$~ case. In Section
4 we present our results on the multiplet classification of the
representations and \idos\ between them.

\newsec{Preliminaries}

\nt Let $G$ be a semisimple non-compact Lie group, and $K$ a maximal
compact subgroup of $G$. Then we have an Iwasawa decomposition
~$G=KA_0N_0$, where ~$A_0$~ is abelian simply connected vector
subgroup of ~$G$, ~$N_0$~ is a nilpotent simply connected subgroup
of ~$G$~ preserved by the action of ~$A_0$. Further, let $M_0$ be
the centralizer of $A_0$ in $K$. Then the subgroup ~$P_0 ~=~ M_0 A_0
N_0$~ is a minimal parabolic subgroup of $G$. A parabolic subgroup
~$P ~=~ M' A' N'$~ is any subgroup of $G$ (including $G$ itself)
which contains a minimal parabolic subgroup.\foot{The number of
non-conjugate parabolic subgroups is ~$2^r$, where $r=\rank\,A$,
cf., e.g., \War.}

The importance of the parabolic subgroups comes from the fact that
the representations induced from them generate all (admissible)
irreducible representations of $G$ \Lan. For the classification of
all irreducible representations it is enough to use only the
so-called {\it cuspidal} parabolic subgroups ~$P=M'A'N'$, singled
out by the condition that ~rank$\, M' =$ rank$\, M'\cap K$
\Zhea,\KnZu, so that $M'$ has discrete series representations \Har.
However, often induction from non-cuspidal parabolics is also
convenient, cf. \EHW,\Dobinv,\Dobpeds.

Let ~$\nu$~ be a (non-unitary) character of ~$A'$, ~$\nu\in\ca'^*$,
let ~$\mu$~ fix an irreducible representation ~$D^\mu$~ of ~$M'$~ on
a vector space ~$V_\mu\,$.

 We call the induced
representation ~$\chi =$ Ind$^G_{P}(\mu\otimes\nu \otimes 1)$~ an
~{\it elementary representation} of $G$ \DMPPT. (These are called
{\it generalized principal series representations} (or {\it limits
thereof}) in \Knapp.) Their spaces of functions are: \eqn\fun{
\cc_\chi ~=~ \{ \cf \in C^\infty(G,V_\mu) ~ \vr ~ \cf (gman) ~=~
e^{-\nu(H)} \cdot D^\mu(m^{-1})\, \cf (g) \} } where ~$a= \exp(H)\in
A'$, ~$H\in\ca'\,$, ~$m\in M'$, ~$n\in N'$.
The representation action is the $left$ regular action:
\eqn\lrr{
(\ct^\chi(g)\cf) (g') ~=~ \cf (g^{-1}g') ~, \quad g,g'\in G\ .}

For our purposes we need to restrict to ~{\it maximal}~ parabolic
subgroups ~$P$, (so that $\rank\,A'=1$), that may not be cuspidal.
For the representations that we consider the character ~$\nu$~ is
parameterized by a real number ~$d$, called the conformal weight or
energy.

Further, let ~$\mu$~ fix a discrete series representation ~$D^\mu$~ of $M'$ on
the Hilbert space ~$V_\mu\,$, or the so-called limit of a discrete
series representation (cf. \Knapp). Actually, instead of the discrete series
we can use the finite-dimensional (non-unitary) representation of $M'$ with the same Casimirs.

An important ingredient in our considerations are the ~{\it
highest/lowest weight representations}~ of ~$\cg$. These can be
realized as (factor-modules of) Verma modules ~$V^\L$~ over
~$\cg^\bac$, where ~$\L\in (\ch^\bac)^*$, ~$\ch^\bac$ is a Cartan
subalgebra of ~$\cg^\bac$, weight ~$\L = \L(\chi)$~ is determined
uniquely from $\chi$ \Dob. In this setting we can consider also
unitarity, which here means positivity w.r.t. the Shapovalov form in
which the conjugation is the one singling out $\cg$ from $\cg^\bac$.

Actually, since our ERs may be induced from finite-dimensional
representations of ~$\cm'$~ (or their limits) the Verma modules are always reducible.
Thus, it is more convenient to use ~{\it generalized Verma modules}
~$\tV^\L$~ such that the role of the highest/lowest weight vector $v_0$ is
taken by the (finite-dimensional) space ~$V_\mu\,v_0\,$. For the
generalized Verma modules (GVMs) the reducibility is controlled only
by the value of the conformal weight $d$.
Relatedly, for the \idos{} only the reducibility w.r.t. non-compact roots is essential.

One main ingredient of our approach is as follows. We group the
(reducible) ERs with the same Casimirs in sets called ~{\it
multiplets} \Dobmul,\Dob{}. The multiplet corresponding to fixed
values of the Casimirs may be depicted as a connected graph, the
vertices of which correspond to the reducible ERs and the lines
between the vertices correspond to intertwining operators.\foot{For
simplicity only the operators which are not compositions of other
operators are depicted.} The explicit parametrization of the
multiplets and of their ERs is important for understanding of the
situation.

In fact, the multiplets contain explicitly all the data necessary to
construct the \idos{}. Actually, the data for each \ido{} consists
of the pair ~$(\b,m)$, where $\b$ is a (non-compact) positive root
of ~$\cg^\bac$, ~$m\in\bbn$, such that the BGG \BGG{} Verma module
reducibility condition (for highest weight modules) is fulfilled:
\eqn\bggr{ (\L+\r, \b^\vee ) ~=~ m \ , \quad \b^\vee \equiv 2 \b
/(\b,\b) \ .} When \bggr{} holds then the Verma module with shifted
weight ~$V^{\L-m\b}$ (or ~$\tV^{\L-m\b}$ ~ for GVM and $\b$
non-compact) is embedded in the Verma module ~$V^{\L}$ (or
~$\tV^{\L}$). This embedding is realized by a singular vector
~$v_s$~ determined by a polynomial ~$\cp_{m,\b}(\cg^-)$~ in the
universal enveloping algebra ~$(U(\cg_-))\ v_0\,$, ~$\cg^-$~ is the
subalgebra of ~$\cg^\bac$ generated by the negative root generators
\Dix.
 More explicitly, \Dob, ~$v^s_{m,\b} = \cp_{m,\b}\, v_0$ (or ~$v^s_{m,\b} = \cp_{m,\b}\, V_\mu\,v_0$ for GVMs).\foot{For
explicit expressions for singular vectors we refer to \Dobsin.} Then
there exists \Dob{} an \ido{} \eqn\lido{\cd_{m,\b} ~:~
\cc_{\chi(\L)} ~\llr ~ \cc_{\chi(\L-m\b)} } given explicitly by:
\eqn\mido{\cd_{m,\b} ~=~ \cp_{m,\b}(\widehat{\cg^-}) } where
~$\widehat{\cg^-}$~ denotes the $right$ action on the functions
~$\cf$, cf. \fun.

\newsec{The non-compact Lie algebra $E_{6(-14)}$}

\nt Let ~$\cg ~=~ E_{6(-14)}\,$, \Helg. This non-compact Lie algebra
is also denoted as EIII \Bourb, or $E'''_6\,$ \Gilm. The maximal
compact subalgebra is ~$\ck \cong so(10)\oplus so(2)$,
~$\dim_\bbr\,\cp = 32$, ~$\dim_\bbr\,\cn^\pm = 30$. This real form
has discrete series representations and highest/lowest weight
representations. The split rank is equal to 2, while ~$\cm_0 \cong
so(6)\oplus so(2)$.

The Satake diagram \Sata{} is: \eqn\satsixay{\underbrace{
\downcirc{{\a_1}} \riga \black{\a_3} \riga
\black{{\a_4}}\kern-8pt\raise11pt\hbox{$\vert$}
\kern-3.5pt\raise22pt\hbox{$\circ {{\scriptstyle{\a_2}}}$}
\riga\black{{\a_5}} \riga\downcirc{{\a_6}} }} Thus, the reduced root
system is presented by a Dynkin-Satake diagram looking like the
~$B_2$~ Dynkin diagram with simple roots ~$\l_1,\l_2\,$, but the
long roots (including $\l_1$) have multiplicity 6, while the short
roots (including $\l_2$) have multiplicity 8, and there are also the
roots $2\l$ of multiplicity 1, where $\l$ is any short root. It is
obtained from \satsixay{} by dropping the black nodes, (they give
rise to $\cm_0$), identifying ~$\a_1$~ and ~$\a_6$~ and mapping them
to ~$\l_2$, while the root ~$\a_2$~ is mapped to ~$\l_1\,$.

The non-minimal parabolic subalgebras are given by: \eqna\parabsixc
$$\eqalignno{
&\cm ~=~ su(5,1)\ , \qquad \dim\,\cn^\pm ~=~ 21\ , \qquad \dim\,\ca
~=~ 1 &\parabsixc a\cr &\cm' ~=~ so(7,1) \oplus so(2)\ , \qquad
\dim\,\cn'^\pm ~=~ 24\ , \qquad \dim\,\ca' ~=~ 1 &\parabsixc b\cr
}$$ Both are maximal and the first is cuspidal.

We shall induce our representations from the first parabolic ~$\cp = \cm \oplus \ca \oplus \cn$.

We label the signature of the ERs of $\cg$ as follows: \eqn\sgnd{
\chi ~~=~~ \{\, n_1\,, n_3\,,n_4\,,n_5\,,n_6\,;\, c \} \ , \quad c =
d- \hel\ , }  where the last entry of ~$\chi$~ labels the characters
of $\ca\,$, and the first five entries are labels of the discrete
series of ~$\cm$, then ~$n_j \in \bbn$, or of limits of discrete
series, when some of $n_j$ are zero. (When ~$n_j \in \bbn$~ the
first five entries label also the finite-dimensional nonunitary
irreps of $\cm\,$, and the finite-dimensional unitary irreps of
~$su(6)$).

The reason to use the parameter ~$c$~ instead of ~$d$~ is that the
parametrization of the ERs in the multiplets is given in a simpler way.

Further, we need the root system of the complex algebra ~$E_6\,$.
With Dynkin diagram enumerating the simple roots ~$\a_i$~ as in
\satsixay, the positive roots are: ~first there are the $15$
 positive roots of ~$sl(6)$~ with~simple~roots~~
$\a_1,\a_3,\a_4,\a_5,\a_6\,$, ~~then there are the following ~21~
roots: \eqna\satsixz
$$\eqalignno{
&\a_2\ , ~~\a_2+\a_4\ , ~~\a_2+\a_4+\a_3\ , ~~\a_2+\a_4+\a_5\
,&\satsixz {}\cr &\a_2+\a_4+\a_3+\a_5\ , ~~\a_2+\a_4+\a_3+\a_1\ ,
~~\a_2+\a_4+\a_5+\a_6\ , \cr& \a_2+\a_4+\a_3+\a_5+\a_1\ ,
~~\a_2+\a_4+\a_3+\a_5+\a_6\ , ~~\a_2+2\a_4+\a_3+\a_5 \ ,\cr&
\a_2+\a_4+\a_3+\a_5+\a_1+\a_6 \ , ~~\a_2+2\a_4+\a_3+\a_5+\a_1\ ,
~~\a_2+2\a_4+\a_3+\a_5+\a_6\ , \cr& \a_2+2\a_4+\a_3+\a_5+\a_1+\a_6\
, ~~\a_2+2\a_4+2\a_3+\a_5+\a_1\ , ~~\a_2+2\a_4+\a_3+2\a_5+\a_6\ ,
\cr& \a_2+2\a_4+2\a_3+\a_5+\a_1+\a_6\ ,
~~\a_2+2\a_4+\a_3+2\a_5+\a_1+\a_6\ , \cr&
\a_2+2\a_4+2\a_3+2\a_5+\a_1+\a_6\ ,\cr&
\a_2+3\a_4+2\a_3+2\a_5+\a_1+\a_6\ , \cr&
2\a_2+3\a_4+2\a_3+2\a_5+\a_1+\a_6 ~\equiv~ \ta \ , }$$ where ~$\ta$~
is the highest root of the $E_6$ root system.

Relative to our parabolic subalgebra, the roots in \satsixz{} are
non-compact, while the rest are compact. The differential
intertwining operators that give the multiplets correspond to the
noncompact roots, and since we shall use the latter extensively, we
introduce more compact notation for them. Namely, the nonsimple
roots in \satsixz{} will be denoted in a self-explanatory way as
follows: \eqna\nota
$$\eqalignno{
&\a_{ij} ~=~ \a_i + \a_{i+1} + \cdots + \a_j \ , ~~~\a_{i,j} ~=~
\a_i + \a_j\ , ~~~i < j \ ,&\nota {} \cr &\a_{ij,k} ~=~ \a_{k,ij}
~=~\a_i + \a_{i+1} +\cdots + \a_j +\a_k\ , ~~~i< j \ , \cr
&\a_{ij,km} ~=~ \a_i + \a_{i+1} +\cdots + \a_j +\a_k+ \a_{k+1}
+\cdots +\a_m\ , ~~~i< j \ , ~~k<m \ , \cr &\a_{ij,km,4} ~=~ \a_i +
\a_{i+1} + \cdots + \a_j +\a_k+ \a_{k+1} + \cdots +\a_m+\a_4\ ,
~~~i< j \ , ~~k<m \ , \cr}$$ i.e., the roots \satsixz{} will be
written as: \eqna\satsixzz
$$\eqalignno{
&\a_2\ , ~~\a_{2,4}\ , ~~\a_{24}\ , ~~\a_{2,45}\ , ~~\a_{25}\ ,
~~\a_{14}\ , ~~\a_{2,46}\ , &\satsixzz {}\cr & \a_{15}\ ,~~\a_{26}\
, ~~\a_{25,4}\ , ~~\a_{16} \ , ~~\a_{15,4} \ ,~~\a_{26,4}\ , \cr
&\a_{16,4} \ , ~~ \a_{15,34} \ ,~~ \a_{26,45} \ , ~~\a_{16,34} \ ,~~
\a_{16,45} \ , \cr &\a_{16,35}\ , ~~ \a_{16,35,4}\ , ~~ \a_{16,25,4}
~=~ \ta \ .
 }$$

Further, we give the correspondence between the signatures $\chi$
and the highest weight $\L$. The connection is through the Dynkin
labels: \eqn\dynk{ m_i ~\equiv~ (\L+\r,\a^\vee_i) ~=~ (\L+\r, \a_i
)\ , \quad i=1,\ldots,6,} where ~$\L = \L(\chi)$, ~$\r$ is half the
sum of the positive roots of ~$\cg^\bac$, ~$\a_i$~ denotes the
simple roots of ~$\cg^\bac$. The explicit connection is: \eqn\rela{
n_i = m_i \ , \quad - c ~=~ \ha n_\ta ~=~ \ha (n_1+2n_2 + 2n_3 + 3n_4 +
2n_5 + n_6) }

We shall use also the so-called Harish-Chandra parameters:
\eqn\dynhc{ m_\b \equiv (\L+\r, \b )\ ,} where $\b$ is any positive
root of $\cg^\bac$. These parameters are redundant, since obviously
they are expressed in terms of the Dynkin labels, however, some
statements are best formulated in their terms.\foot{Clearly, both
the Dynkin and Harish-Chandra labels have their origin in the BGG
reducibility condition \bggr.}

There are several types of multiplets: the main type, which contains
maximal number of ERs/GVMs, the finite-dimensional and the discrete
series representations, and some reduced types of multiplets.

In the next Section we give the classification of all multiplets that are
physically relevant.

\newsec{Multiplets}

\subsec{Main type of multiplets}

\nt The multiplets of the main type are in 1-to-1 correspondence
with the finite-dimensional irreps of ~$\cg\,$, i.e., they will be
labelled by the six positive Dynkin labels ~$m_i\in\bbn$. It turns
out that each such multiplet contains 70 ERs/GVMs whose signatures
can be given in the following pair-wise manner: \eqna\tabl
$$\eqalignno{
&\chi_0^\pm ~=~ \{\, m_1\,, m_3\,,m_4\,,m_5\,,m_6\,;\, \pm\ha
m_\ta\, \} \ , &\tabl{}\cr &\chi_a^\pm ~=~ \{\, m_1\,,
m_3\,,m_{2,4}\,,m_5\,,m_6\,;\, \pm\ha (m_\ta-m_2)\, \} \ , \cr
&\chi_b^\pm ~=~ \{\, m_1\,, m_{34}\,,m_{2}\,,m_{45}\,,m_6\,;\,
\pm\ha (m_\ta-m_{2,4})\, \} \ , \cr &\chi_c^\pm ~=~ \{\, m_{1,3}\,,
m_{4}\,,m_{2}\,,m_{35}\,,m_6\,;\, \pm\ha (m_\ta-m_{24})\, \} \ , \cr
&\chi_{c'}^\pm ~=~ \{\, m_{1}\,,
m_{35}\,,m_{2}\,,m_{4}\,,m_{56}\,;\, \pm\ha (m_\ta-m_{2,45})\, \} \
, \cr &\chi_{d}^\pm ~=~ \{\, m_{3}\,,
m_{4}\,,m_{2}\,,m_{1,35}\,,m_{6}\,;\, \pm\ha (m_\ta-m_{14})\, \} \ ,
\cr &\chi_{d'}^\pm ~=~ \{\, m_{1}\,,
m_{36}\,,m_{2}\,,m_{4}\,,m_{5}\,;\, \pm\ha (m_\ta-m_{2,46})\, \} \ ,
\cr &\chi_{\td}^\pm ~=~ \{\, m_{1,3}\,,
m_{45}\,,m_{2}\,,m_{34}\,,m_{56}\,;\, \pm\ha (m_\ta-m_{25})\, \} \ ,
\cr &\chi_{e}^\pm ~=~ \{\, m_{3}\,,
m_{45}\,,m_{2}\,,m_{1,34}\,,m_{56}\,;\, \pm\ha (m_\ta-m_{15})\, \} \
, \cr &\chi_{e'}^\pm ~=~ \{\, m_{1,3}\,,
m_{46}\,,m_{2}\,,m_{34}\,,m_{5}\,;\, \pm\ha (m_\ta-m_{26})\, \} \ ,
\cr &\chi_{\te}^\pm ~=~ \{\, m_{3}\,,
m_{46}\,,m_{2}\,,m_{1,34}\,,m_{5}\,;\, \pm\ha (m_\ta-m_{16})\, \} \
, \cr &\chi_{f}^\pm ~=~ \{\, m_{34}\,,
m_{5}\,,m_{2,4}\,,m_{1,3}\,,m_{46}\,;\, \pm\ha (m_\ta-m_{15,4})\, \}
\ , \cr &\chi_{f'}^\pm ~=~ \{\, m_{1,34}\,,
m_{56}\,,m_{2,4}\,,m_{3}\,,m_{45}\,;\, \pm\ha (m_\ta-m_{26,4})\, \}
\ , \cr &\chi_{\tf}^\pm ~=~ \{\, m_{1,34}\,,
m_{5}\,,m_{2,4}\,,m_{3}\,,m_{46}\,;\, \pm\ha (m_\ta-m_{25,4})\, \} \
, \cr &\chi_{f^o}^\pm ~=~ \{\, m_{4}\,,
m_{5}\,,m_{24}\,,m_{1}\,,m_{36}\,;\, \pm\ha (m_\ta-m_{15,34})\, \} \
, \cr &\chi_{f''}^\pm ~=~ \{\, m_{1,35}\,,
m_{6}\,,m_{2,45}\,,m_{3}\,,m_{4}\,;\, \pm\ha (m_\ta-m_{26,45})\, \}
\ , \cr &\chi_{g}^\pm ~=~ \{\, m_{24}\,,
m_{5}\,,m_{4}\,,m_{1,3}\,,m_{2,46}\,;\, \pm\ha m_{36}\, \} \ , \cr
&\chi_{g'}^\pm ~=~ \{\, m_{14}\,,
m_{56}\,,m_{4}\,,m_{3}\,,m_{2,45}\,;\,\pm\ha m_{1,35}\, \} \ , \cr
&\chi_{\tg}^\pm ~=~ \{\, m_{14}\,,
m_{5}\,,m_{4}\,,m_{3}\,,m_{2,46}\,;\, \pm\ha m_{1,36}\, \} \ , \cr
&\chi_{g^o}^\pm ~=~ \{\, m_{2,4}\,,
m_{5}\,,m_{34}\,,m_{1}\,,m_{26}\,;\, \pm\ha m_{46}\, \} \ , \cr
&\chi_{g''}^\pm ~=~ \{\, m_{15}\,,
m_{6}\,,m_{45}\,,m_{3}\,,m_{2,4}\,;\, \pm\ha m_{1,34}\, \} \ , \cr
&\chi_{h}^\pm ~=~ \{\, m_{4}\,,
m_{56}\,,m_{24}\,,m_{1}\,,m_{35}\,;\,\pm\ha m_{2,45}\, \} \ , \cr
&\chi_{h'}^\pm ~=~ \{\, m_{35}\,,
m_{6}\,,m_{2,45}\,,m_{1,3}\,,m_{4}\,;\, \pm\ha m_{24}\, \} \ , \cr
&\chi_{\th}^\pm ~=~ \{\, m_{34}\,,
m_{56}\,,m_{2,4}\,,m_{1,3}\,,m_{45}\,;\,\pm\ha m_{25}\, \} \ , \cr
&\chi_{\hh}^\pm ~=~ \{\, m_{45}\,,
m_{6}\,,m_{25}\,,m_{1}\,,m_{34}\,;\, \pm\ha m_{2,4}\, \} \ , \cr
&\chi_{\hk}^\pm ~=~ \{\, m_{24}\,,
m_{56}\,,m_{4}\,,m_{1,3}\,,m_{2,45}\,;\, \pm\ha m_{35}\, \} \ , \cr
&\chi_{j}^\pm ~=~ \{\, m_{2,4}\,,
m_{56}\,,m_{34}\,,m_{1}\,,m_{25}\,;\, \pm\ha m_{45}\, \} \ , \cr
&\chi_{j'}^\pm ~=~ \{\, m_{25}\,,
m_{6}\,,m_{45}\,,m_{1,3}\,,m_{2,4}\,;\, \pm\ha m_{34}\, \} \ , \cr
&\chi_{j^o}^\pm ~=~ \{\, m_{2}\,,
m_{45}\,,m_{3}\,,m_{1}\,,m_{26,4}\,;\, \pm\ha m_{56}\, \} \ , \cr
&\chi_{j''}^\pm ~=~ \{\, m_{15,4}\,,
m_{6}\,,m_{5}\,,m_{34}\,,m_{2}\,;\, \pm\ha m_{1,3}\, \} \ , \cr
&\chi_{k}^\pm ~=~ \{\, m_{2}\,,
m_{46}\,,m_{3}\,,m_{1}\,,m_{25,4}\,;\, \pm\ha m_{5}\, \} \ , \cr
&\chi_{k'}^\pm ~=~ \{\, m_{25,4}\,,
m_{6}\,,m_{5}\,,m_{1,34}\,,m_{2}\,;\, \pm\ha m_{3}\, \} \ , \cr
&\chi_{\tk}^\pm ~=~ \{\, m_{2,45}\,,
m_{6}\,,m_{35}\,,m_{1}\,,m_{24}\,;\, \pm\ha m_{4}\, \} \ , \cr
&\chi_{k^o}^\pm ~=~ \{\, m_{2}\,,
m_{4}\,,m_{3}\,,m_{1}\,,m_{26,45}\,;\, \pm\ha m_{6}\, \} \ , \cr
&\chi_{k''}^\pm ~=~ \{\, m_{15,34}\,,
m_{6}\,,m_{5}\,,m_{4}\,,m_{2}\,;\, \pm\ha m_{1}\, \} \ , \cr }$$
where we have used for the numbers ~$m_\b ~=~ (\L(\chi)+\r,\b)$~ the
same compact notation as in \nota{} for the roots $\b$.

The ERs in the multiplet are related by intertwining integral and
differential operators. The integral operators were introduced by
Knapp and Stein \KnSt{}. The relevant fact here is that the above
pairs are related by Knapp-Stein integral operators, such that the
weights are reflected by the maximal root: \eqn\weylr{ s_\ta \cdot
\L(\chi^\pm) = \L(\chi^\mp) } These operators intertwining the pairs
will be denoted by: \eqn\ackin{ G^\pm ~:~ \cc_{\chi^\mp} \lra
\cc_{\chi^\pm} }

Matters are arranged so that in every multiplet only the ER with
signature ~$\chi_0^-$~ contains a finite-dimensional nonunitary
subrepresentation in a finite-dimensional subspace ~$\ce$. The
latter corresponds to the finite-dimensional irrep of ~$E_6$~ with
signature ~$\{ m_1\,, \ldots,\, m_6 \}$. The subspace ~$\ce$~ is
annihilated by the operator ~$G^+\,$,\ and is the image of the
operator ~$G^-\,$. When all ~$m_i=1$~ then ~$\dim\,\ce = 1$, and in
that case ~$\ce$~ is also the trivial one-dimensional UIR of the
whole algebra ~$E_{6(-14)}$. Furthermore in that case the conformal
weight is zero: ~$d=\hel+c=\ha (11-{m_{\ta}}_{\vert_{m_i=1}})=0$.

Analogously, in every multiplet only the ER with signature
~$\chi_0^+$~ contains holomorphic discrete series representation.
This is guaranteed by the fact that for this ER all Harish-Chandra
parameters for non-compact roots are negative, i.e., ~$ n_\a ~<~ 0$,
for ~$\a$~ from \satsixz{}. [The last fact can be easily checked
from the signatures \tabl{}.] The conformal weight has the
restriction ~$d = \ha (11 + m_\ta) \geq 11$.

In fact, this ER contains also the conjugate anti-holomorphic
discrete series. The direct sum of the holomorphic and the
antiholomorphic representations are realized in an invariant
subspace ~$\cd$~ of the ER ~$\chi_0^+\,$. That subspace is
annihilated by the operator ~$G^-\,$,\ and is the image of the
operator ~$G^+\,$.

Note that the corresponding lowest weight GVM is infinitesimally
equivalent only to the holomorphic discrete series, while the
conjugate highest weight GVM is infinitesimally equivalent to the
anti-holomorphic discrete series.

The \idos\ correspond to non-compact positive roots of the root
system of ~$E_6$, cf. \Dob, i.e., in the current context, the roots
 given in \satsixz{}, (or in more
compact notation, \satsixzz{}).

The multiplets are given explicitly in Fig. 1. Each \ido\ is
represented by an arrow accompanied by a symbol ~$i_{j...k}$~
encoding the root ~$\b_{j...k}$~ and the number $m_{\b_{j...k}}$
which is involved in the BGG criterion. This notation is used to
save space, but it can be used due to
 the fact that only \idos\ which are
non-composite are displayed, and that the data ~$\b,m_\b\,$, which
is involved in the embedding ~$V^\L \lra V^{\L-m_\b,\b}$~ turns out
to involve only the ~$m_i$~ corresponding to simple roots, i.e., for
each $\b,m_\b$ there exists ~$i = i(\b,m_\b,\L)\in \{ 1,\ldots,6\}$,
such that ~$m_\b=m_i\,$. Hence the data ~$\b_{j...k}\,$,$m_{\b_{j...k}}$~
is represented by ~$i_{j...k}$~ on the arrows.

Note that there are five cases when the embeddings correspond to the
highest root $\ta$~: ~~$V^{\L^-} \lra V^{\L^+}$, ~$\L^+ ~=~\L^-
-m_\ta\,\ta\,$. In these five cases the weights are denoted as:
~$\L^\pm_{k''}\,$, ~$\L^\pm_{k'}\,$, ~$\L^\pm_{\tk}\,$,
~$\L^\pm_{k}\,$, ~$\L^\pm_{k^o}\,$, then we have: ~$m_\ta ~=~
m_1,m_3,m_4,m_5,m_6\,$, resp. We recall that Knapp-Stein operators
~$G^+$~ intertwine the corresponding ERs ~$\ct_\chi^-$ and
~$\ct_\chi^+$, cf. \ackin{}. In the above five cases the Knapp-Stein
operators ~$G^+$~ degenerate to differential operators.
 The latter phenomenon is given explicitly for the case of the anti-de-Sitter
 algebra $so(3,2)$ in \Dobso.

Note that the figure has the standard $E_6$ symmetry, namely, conjugation exchanging indices $1\llra 6$, $3\llra 5$.
This conjugation will be used also below when we describe the reduced embedding diagrams.

In the next Subsections we shall consider the reduced types of multiplets.

\subsec{Main types of reduced multiplets}
\subsubsecno=0

\subsubsec{Reduced type R2 multiplets}

\nt
The multiplets of reduced type R2 contain 50 ERs/GVMs and may be obtained formally from the main type by
setting ~$m_2 ~=~ 0$. Their signatures are given explicitly by:
\eqna\tabla
$$\eqalignno{
&\chi_a^\pm ~=~ \{\, m_1\,, m_3\,,m_4\,,m_5\,,m_6\,;\, \pm\ha
m_\ta\, \} \ , &\tabla{}\cr &\chi_b^\pm ~=~ \{\, m_1\,,
m_{34}\,,0\,,m_{45}\,,m_6\,;\, \pm\ha (m_\ta-m_{4})\, \} \ , \cr
&\chi_c^\pm ~=~ \{\, m_{1,3}\,, m_{4}\,,0\,,m_{35}\,,m_6\,;\, \pm\ha
(m_\ta-m_{34})\, \} \ , \cr &\chi_{c'}^\pm ~=~ \{\,
m_{1}\,,m_{35}\,,0\,,m_{4}\,,m_{56}\,;\, \pm\ha (m_\ta-m_{45})\, \}
\ , \cr &\chi_{d}^\pm ~=~ \{\,
m_{3}\,,m_{4}\,,0\,,m_{1,35}\,,m_{6}\,;\, \pm\ha (m_\ta-m_{1,34})\,
\} \ , \cr &\chi_{d'}^\pm ~=~ \{\,
m_{1}\,,m_{36}\,,0\,,m_{4}\,,m_{5}\,;\, \pm\ha (m_\ta-m_{46})\, \} \
, \cr &\chi_{\td}^\pm ~=~ \{\,
m_{1,3}\,,m_{45}\,,0\,,m_{34}\,,m_{56}\,;\, \pm\ha (m_\ta-m_{35})\,
\} \ , \cr &\chi_{e}^\pm ~=~ \{\,
m_{3}\,,m_{45}\,,0\,,m_{1,34}\,,m_{56}\,;\, \pm\ha
(m_\ta-m_{1,35})\, \} \ , \cr &\chi_{e'}^\pm ~=~ \{\,
m_{1,3}\,,m_{46}\,,0\,,m_{34}\,,m_{5}\,;\, \pm\ha (m_\ta-m_{36})\,
\} \ , \cr &\chi_{\te}^\pm ~=~ \{\,
m_{3}\,,m_{46}\,,0\,,m_{1,34}\,,m_{5}\,;\, \pm\ha (m_\ta-m_{1,36})\,
\} \ , \cr &\chi_{g}^\pm ~=~ \{\,
m_{34}\,,m_{5}\,,m_{4}\,,m_{1,3}\,,m_{46}\,;\, \pm\ha m_{36}\, \} \
, \cr &\chi_{g'}^\pm ~=~ \{\,
m_{1,34}\,,m_{56}\,,m_{4}\,,m_{3}\,,m_{45}\,;\,\pm\ha m_{1,35}\, \}
\ , \cr &\chi_{\tg}^\pm ~=~ \{\,
m_{1,34}\,,m_{5}\,,m_{4}\,,m_{3}\,,m_{46}\,;\, \pm\ha m_{1,36}\, \}
\ , \cr &\chi_{g^o}^\pm ~=~ \{\,
m_{4}\,,m_{5}\,,m_{34}\,,m_{1}\,,m_{36}\,;\, \pm\ha m_{46}\, \} \ ,
\cr &\chi_{g''}^\pm ~=~ \{\,
m_{1,35}\,,m_{6}\,,m_{45}\,,m_{3}\,,m_{4}\,;\, \pm\ha m_{1,34}\, \}
\ , \cr &\chi_{\hk}^\pm ~=~ \{\,
m_{34}\,,m_{56}\,,m_{4}\,,m_{1,3}\,,m_{45}\,;\, \pm\ha m_{35}\, \} \
, \cr &\chi_{j}^\pm ~=~ \{\,
m_{4}\,,m_{56}\,,m_{34}\,,m_{1}\,,m_{35}\,;\, \pm\ha m_{45}\, \} \ ,
\cr &\chi_{j'}^\pm ~=~ \{\,
m_{35}\,,m_{6}\,,m_{45}\,,m_{1,3}\,,m_{4}\,;\, \pm\ha m_{34}\, \} \
, \cr &\chi_{j^o}^\pm ~=~ \{\,
0\,,m_{45}\,,m_{3}\,,m_{1}\,,m_{36,4}\,;\, \pm\ha m_{56}\, \} \ ,
\cr &\chi_{j''}^\pm ~=~ \{\,
m_{1,35,4}\,,m_{6}\,,m_{5}\,,m_{34}\,,0\,;\, \pm\ha m_{1,3}\, \} \ ,
\cr &\chi_{k}^\pm ~=~ \{\,
0\,,m_{46}\,,m_{3}\,,m_{1}\,,m_{35,4}\,;\, \pm\ha m_{5}\, \} \ , \cr
&\chi_{k'}^\pm ~=~ \{\, m_{35,4}\,,m_{6}\,,m_{5}\,,m_{1,34}\,,0\,;\,
\pm\ha m_{3}\, \} \ , \cr &\chi_{\tk}^\pm ~=~ \{\,
m_{45}\,,m_{6}\,,m_{35}\,,m_{1}\,,m_{34}\,;\, \pm\ha m_{4}\, \} \ ,
\cr &\chi_{k^o}^\pm ~=~ \{\,
0\,,m_{4}\,,m_{3}\,,m_{1}\,,m_{36,45}\,;\, \pm\ha m_{6}\, \} \ , \cr
&\chi_{k''}^\pm ~=~ \{\,
m_{1,35,34}\,,m_{6}\,,m_{5}\,,m_{4}\,,0\,;\, \pm\ha m_{1}\, \} \ \cr
}$$ These multiplets are depicted on Fig. 2. To save space we give
only the ~$\L^-(\chi)$~ modules, since (as we know from the Main
type, Fig. 1.), the ~$\L^+(\chi)$~ part of the multiplets has the
same structure.

Here the ER ~$\chi_0^+$~ contains limits of the (anti)holomorphic discrete series representations.
This is guaranteed by the fact that for this ER
all Harish-Chandra parameters for non-compact roots are non-positive, i.e.,
~$ n_\a ~\leq~ 0$, for ~$\a$~ from \satsixz{}.
The conformal weight has the restriction ~$d = \ha (11 + m_\ta) \geq 10$.

There are other limiting cases, where there are zero entries for the first five ~$n_i$~ values.
In these cases the induction procedure would not use finite-dimensional irreps of the ~$su(5,1)$~
subgroup. The corresponding ERs would not have direct physical meaning,
however, the fact that they are together with the physically meaningful ERs has important
bearing on the structure of the latter.

\subsubsec{Reduced type R3 multiplets}

\nt The multiplets of reduced type R3 contain 49 ERs/GVMs and may be
obtained formally from the main type by setting ~$m_3 ~=~ 0$. Their
signatures are given explicitly by: \eqna\tablb
$$\eqalignno{
&\chi_0^\pm ~=~ \{\, m_1\,, 0\,,m_4\,,m_5\,,m_6\,;\, \pm\ha m_\ta\,
\} \ , &\tablb{}\cr &\chi_a^\pm ~=~ \{\, m_1\,,
0\,,m_{2,4}\,,m_5\,,m_6\,;\, \pm\ha (m_\ta-m_2)\, \} \ , \cr
 &\chi_c^\pm ~=~ \{\, m_{1}\,,
m_{4}\,,m_{2}\,,m_{45}\,,m_6\,;\, \pm\ha (m_\ta-m_{2,4})\, \} \ ,
\cr &\chi_{d}^\pm ~=~ \{\, 0\,,
m_{4}\,,m_{2}\,,m_{1,45}\,,m_{6}\,;\, \pm\ha (m_\ta-m_{12,4})\, \} \
, \cr &\chi_{\td}^\pm ~=~ \{\, m_{1}\,,
m_{45}\,,m_{2}\,,m_{4}\,,m_{56}\,;\, \pm\ha (m_\ta-m_{2,45})\, \} \
, \cr &\chi_{e}^\pm ~=~ \{\, 0\,,
m_{45}\,,m_{2}\,,m_{1,4}\,,m_{56}\,;\, \pm\ha (m_\ta-m_{12,45})\, \}
\ , \cr &\chi_{e'}^\pm ~=~ \{\, m_{1}\,,
m_{46}\,,m_{2}\,,m_{4}\,,m_{5}\,;\, \pm\ha (m_\ta-m_{2,46})\, \} \ ,
\cr &\chi_{\te}^\pm ~=~ \{\, 0\,,
m_{46}\,,m_{2}\,,m_{1,4}\,,m_{5}\,;\, \pm\ha (m_\ta-m_{12,46})\, \}
\ , \cr &\chi_{f}^\pm ~=~ \{\, m_{4}\,,
m_{5}\,,m_{2,4}\,,m_{1}\,,m_{46}\,;\, \pm\ha (m_\ta-m_{12,45,4})\,
\} \ , \cr &\chi_{f'}^\pm ~=~ \{\, m_{1,4}\,,
m_{56}\,,m_{2,4}\,,0\,,m_{45}\,;\, \pm\ha (m_\ta-m_{2,46,4})\, \} \
, \cr &\chi_{\tf}^\pm ~=~ \{\, m_{1,4}\,,
m_{5}\,,m_{2,4}\,,0\,,m_{46}\,;\, \pm\ha (m_\ta-m_{2,45,4})\, \} \ ,
\cr &\chi_{f''}^\pm ~=~ \{\, m_{1,45}\,,
m_{6}\,,m_{2,45}\,,0\,,m_{4}\,;\, \pm\ha (m_\ta-m_{2,46,45})\, \} \
, \cr &\chi_{g}^\pm ~=~ \{\, m_{2,4}\,,
m_{5}\,,m_{4}\,,m_{1}\,,m_{2,46}\,;\, \pm\ha m_{46}\, \} \ , \cr
&\chi_{g'}^\pm ~=~ \{\, m_{12,4}\,,
m_{56}\,,m_{4}\,,0\,,m_{2,45}\,;\,\pm\ha m_{1,45}\, \} \ , \cr
&\chi_{\tg}^\pm ~=~ \{\, m_{12,4}\,,
m_{5}\,,m_{4}\,,0\,,m_{2,46}\,;\, \pm\ha m_{1,46}\, \} \ , \cr
&\chi_{g''}^\pm ~=~ \{\, m_{12,45}\,,
m_{6}\,,m_{45}\,,0\,,m_{2,4}\,;\, \pm\ha m_{1,4}\, \} \ , \cr
&\chi_{h'}^\pm ~=~ \{\, m_{45}\,,
m_{6}\,,m_{2,45}\,,m_{1}\,,m_{4}\,;\, \pm\ha m_{2,4}\, \} \ , \cr
&\chi_{\th}^\pm ~=~ \{\, m_{4}\,,
m_{56}\,,m_{2,4}\,,m_{1}\,,m_{45}\,;\,\pm\ha m_{2,45}\, \} \ , \cr
&\chi_{\hk}^\pm ~=~ \{\, m_{2,4}\,,
m_{56}\,,m_{4}\,,m_{1}\,,m_{2,45}\,;\, \pm\ha m_{45}\, \} \ , \cr
&\chi_{j^o}^\pm ~=~ \{\, m_{2}\,,
m_{45}\,,0\,,m_{1}\,,m_{2,46,4}\,;\, \pm\ha m_{56}\, \} \ , \cr
&\chi_{k}^\pm ~=~ \{\, m_{2}\,, m_{46}\,,0\,,m_{1}\,,m_{2,45,4}\,;\,
\pm\ha m_{5}\, \} \ , \cr &\chi_{k'}^\pm ~=~ \{\, m_{2,45,4}\,,
m_{6}\,,m_{5}\,,m_{1,4}\,,m_{2}\,;\, 0\, \} \ , \cr &\chi_{\tk}^\pm
~=~ \{\, m_{2,45}\,, m_{6}\,,m_{45}\,,m_{1}\,,m_{2,4}\,;\, \pm\ha
m_{4}\, \} \ , \cr &\chi_{k^o}^\pm ~=~ \{\, m_{2}\,,
m_{4}\,,0\,,m_{1}\,,m_{2,46,45}\,;\, \pm\ha m_{6}\, \} \ , \cr
&\chi_{k''}^\pm ~=~ \{\, m_{12,45,4}\,,
m_{6}\,,m_{5}\,,m_{4}\,,m_{2}\,;\, \pm\ha m_{1}\, \} \ \cr }$$ In
these multiplets the number of members is odd, since the ERs
$\chi_{k'}^\pm$ coincide. These multiplets are depicted on Fig. 3.

\subsubsec{Reduced type R4 multiplets}

\nt
The multiplets of reduced type R4 contain 51 ERs/GVMs and may be obtained formally from the main type by
setting ~$m_4 ~=~ 0$. Their signatures are given explicitly by:
\eqna\tablc
$$\eqalignno{
&\chi_0^\pm ~=~ \{\, m_1\,, m_3\,,0\,,m_5\,,m_6\,;\, \pm\ha m_\ta\,
\} \ , &\tablc{}\cr &\chi_b^\pm ~=~ \{\, m_1\,,
m_{3}\,,m_{2}\,,m_{5}\,,m_6\,;\, \pm\ha (m_\ta-m_{2})\, \} \ , \cr
&\chi_c^\pm ~=~ \{\, m_{1,3}\,, 0\,,m_{2}\,,m_{3,5}\,,m_6\,;\,
\pm\ha (m_\ta-m_{23})\, \} \ , \cr &\chi_{c'}^\pm ~=~ \{\, m_{1}\,,
m_{3,5}\,,m_{2}\,,0\,,m_{56}\,;\, \pm\ha (m_\ta-m_{2,5})\, \} \ ,
\cr &\chi_{d}^\pm ~=~ \{\, m_{3}\,,
0\,,m_{2}\,,m_{1,3,5}\,,m_{6}\,;\, \pm\ha (m_\ta-m_{13})\, \} \ ,
\cr &\chi_{d'}^\pm ~=~ \{\, m_{1}\,,
m_{3,56}\,,m_{2}\,,0\,,m_{5}\,;\, \pm\ha (m_\ta-m_{2,56})\, \} \ ,
\cr &\chi_{f}^\pm ~=~ \{\, m_{3}\,,
m_{5}\,,m_{2}\,,m_{1,3}\,,m_{56}\,;\, \pm\ha (m_\ta-m_{13,5})\, \} \
, \cr &\chi_{f'}^\pm ~=~ \{\, m_{1,3}\,,
m_{56}\,,m_{2}\,,m_{3}\,,m_{5}\,;\, \pm\ha (m_\ta-m_{23,56})\, \} \
, \cr &\chi_{\tf}^\pm ~=~ \{\, m_{1,3}\,,
m_{5}\,,m_{2}\,,m_{3}\,,m_{56}\,;\, \pm\ha (m_\ta-m_{23,5})\, \} \ ,
\cr &\chi_{f^o}^\pm ~=~ \{\, 0\,,
m_{5}\,,m_{23}\,,m_{1}\,,m_{3,56}\,;\, \pm\ha (m_\ta-m_{13,5,3})\,
\} \ , \cr &\chi_{f''}^\pm ~=~ \{\, m_{1,3,5}\,,
m_{6}\,,m_{2,5}\,,m_{3}\,,0\,;\, \pm\ha (m_\ta-m_{23,56,5})\, \} \ ,
\cr &\chi_{g}^\pm ~=~ \{\, m_{23}\,,
m_{5}\,,0\,,m_{1,3}\,,m_{2,56}\,;\, \pm\ha m_{3,56}\, \} \ , \cr
&\chi_{g'}^\pm ~=~ \{\, m_{13}\,,
m_{56}\,,0\,,m_{3}\,,m_{2,5}\,;\,\pm\ha m_{1,3,5}\, \} \ , \cr
&\chi_{\tg}^\pm ~=~ \{\, m_{13}\,, m_{5}\,,0\,,m_{3}\,,m_{2,56}\,;\,
\pm\ha m_{1,3,56}\, \} \ , \cr &\chi_{g^o}^\pm ~=~ \{\, m_{2}\,,
m_{5}\,,m_{3}\,,m_{1}\,,m_{23,56}\,;\, \pm\ha m_{56}\, \} \ , \cr
&\chi_{g''}^\pm ~=~ \{\, m_{13,5}\,,
m_{6}\,,m_{5}\,,m_{3}\,,m_{2}\,;\, \pm\ha m_{1,3}\, \} \ , \cr
&\chi_{h}^\pm ~=~ \{\, 0\,,
m_{56}\,,m_{23}\,,m_{1}\,,m_{3,5}\,;\,\pm\ha m_{2,5}\, \} \ , \cr
&\chi_{h'}^\pm ~=~ \{\, m_{3,5}\,,
m_{6}\,,m_{2,5}\,,m_{1,3}\,,0\,;\, \pm\ha m_{23}\, \} \ , \cr
&\chi_{\th}^\pm ~=~ \{\, m_{3}\,,
m_{56}\,,m_{2}\,,m_{1,3}\,,m_{5}\,;\,\pm\ha m_{23,5}\, \} \ , \cr
&\chi_{\hh}^\pm ~=~ \{\, m_{5}\,,
m_{6}\,,m_{23,5}\,,m_{1}\,,m_{3}\,;\, \pm\ha m_{2}\, \} \ , \cr
&\chi_{\hk}^\pm ~=~ \{\, m_{23}\,,
m_{56}\,,0\,,m_{1,3}\,,m_{2,3,5}\,;\, \pm\ha m_{3,5}\, \} \ , \cr
&\chi_{k}^\pm ~=~ \{\, m_{2}\,,
m_{56}\,,m_{3}\,,m_{1}\,,m_{23,5}\,;\, \pm\ha m_{5}\, \} \ , \cr
&\chi_{k'}^\pm ~=~ \{\, m_{23,5}\,,
m_{6}\,,m_{5}\,,m_{1,3}\,,m_{2}\,;\, \pm\ha m_{3}\, \} \ , \cr
&\chi_{\tk}^\pm ~=~ \{\, m_{2,5}\,,
m_{6}\,,m_{3,5}\,,m_{1}\,,m_{23}\,;\, 0\, \} \ , \cr &\chi_{k^o}^\pm
~=~ \{\, m_{2}\,, 0\,,m_{3}\,,m_{1}\,,m_{23,56,5}\,;\, \pm\ha
m_{6}\, \} \ , \cr &\chi_{k''}^\pm ~=~ \{\, m_{13,5,3}\,,
m_{6}\,,m_{5}\,,0\,,m_{2}\,;\, \pm\ha m_{1}\, \} \ \cr }$$ These
multiplets are depicted on Fig. 4.

\subsubsec{Reduced type R1 multiplets}

\nt The multiplets of reduced type R1 contain 49 ERs/GVMs and may be
obtained formally from the main type by setting ~$m_1 ~=~ 0$. Their
signatures are given explicitly by: \eqna\tabld
$$\eqalignno{
&\chi_0^\pm ~=~ \{\, 0\,, m_3\,,m_4\,,m_5\,,m_6\,;\, \pm\ha m_\ta\,
\} \ , &\tabld{}\cr &\chi_a^\pm ~=~ \{\, 0\,,
m_3\,,m_{2,4}\,,m_5\,,m_6\,;\, \pm\ha (m_\ta-m_2)\, \} \ ,\cr
&\chi_b^\pm ~=~ \{\, 0\,, m_{34}\,,m_{2}\,,m_{45}\,,m_6\,;\, \pm\ha
(m_\ta-m_{2,4})\, \} \ , \cr &\chi_{c'}^\pm ~=~ \{\, 0\,,
m_{35}\,,m_{2}\,,m_{4}\,,m_{56}\,;\, \pm\ha (m_\ta-m_{2,45})\, \} \
, \cr
 &\chi_{d}^\pm ~=~ \{\, m_{3}\,,
m_{4}\,,m_{2}\,,m_{35}\,,m_{6}\,;\, \pm\ha (m_\ta-m_{24})\, \} \ ,
\cr &\chi_{d'}^\pm ~=~ \{\, 0\,, m_{36}\,,m_{2}\,,m_{4}\,,m_{5}\,;\,
\pm\ha (m_\ta-m_{2,46})\, \} \ , \cr
 &\chi_{e}^\pm ~=~ \{\, m_{3}\,,
m_{45}\,,m_{2}\,,m_{34}\,,m_{56}\,;\, \pm\ha (m_\ta-m_{25})\, \} \ ,
\cr &\chi_{\te}^\pm ~=~ \{\, m_{3}\,,
m_{46}\,,m_{2}\,,m_{34}\,,m_{5}\,;\, \pm\ha (m_\ta-m_{26})\, \} \ ,
\cr &\chi_{f}^\pm ~=~ \{\, m_{34}\,,
m_{5}\,,m_{2,4}\,,m_{3}\,,m_{46}\,;\, \pm\ha (m_\ta-m_{25,4})\, \} \
, \cr &\chi_{f^o}^\pm ~=~ \{\, m_{4}\,,
m_{5}\,,m_{24}\,,0\,,m_{36}\,;\, \pm\ha (m_\ta-m_{25,34})\, \} \ ,
\cr &\chi_{g}^\pm ~=~ \{\, m_{24}\,,
m_{5}\,,m_{4}\,,m_{3}\,,m_{2,46}\,;\, \pm\ha m_{36}\, \} \ , \cr
&\chi_{g^o}^\pm ~=~ \{\, m_{2,4}\,, m_{5}\,,m_{34}\,,0\,,m_{26}\,;\,
\pm\ha m_{46}\, \} \ , \cr &\chi_{h}^\pm ~=~ \{\, m_{4}\,,
m_{56}\,,m_{24}\,,0\,,m_{35}\,;\,\pm\ha m_{2,45}\, \} \ , \cr
&\chi_{h'}^\pm ~=~ \{\, m_{35}\,,
m_{6}\,,m_{2,45}\,,m_{3}\,,m_{4}\,;\, \pm\ha m_{24}\, \} \ , \cr
&\chi_{\th}^\pm ~=~ \{\, m_{34}\,,
m_{56}\,,m_{2,4}\,,m_{3}\,,m_{45}\,;\,\pm\ha m_{25}\, \} \ , \cr
&\chi_{\hh}^\pm ~=~ \{\, m_{45}\,, m_{6}\,,m_{25}\,,0\,,m_{34}\,;\,
\pm\ha m_{2,4}\, \} \ , \cr &\chi_{\hk}^\pm ~=~ \{\, m_{24}\,,
m_{56}\,,m_{4}\,,m_{3}\,,m_{2,45}\,;\, \pm\ha m_{35}\, \} \ , \cr
&\chi_{j}^\pm ~=~ \{\, m_{2,4}\,, m_{56}\,,m_{34}\,,0\,,m_{25}\,;\,
\pm\ha m_{45}\, \} \ , \cr &\chi_{j'}^\pm ~=~ \{\, m_{25}\,,
m_{6}\,,m_{45}\,,m_{3}\,,m_{2,4}\,;\, \pm\ha m_{34}\, \} \ , \cr
&\chi_{j^o}^\pm ~=~ \{\, m_{2}\,, m_{45}\,,m_{3}\,,0\,,m_{26,4}\,;\,
\pm\ha m_{56}\, \} \ , \cr &\chi_{k}^\pm ~=~ \{\, m_{2}\,,
m_{46}\,,m_{3}\,,0\,,m_{25,4}\,;\, \pm\ha m_{5}\, \} \ , \cr
&\chi_{k'}^\pm ~=~ \{\, m_{25,4}\,,
m_{6}\,,m_{5}\,,m_{34}\,,m_{2}\,;\, \pm\ha m_{3}\, \} \ , \cr
&\chi_{\tk}^\pm ~=~ \{\, m_{2,45}\,,
m_{6}\,,m_{35}\,,0\,,m_{24}\,;\, \pm\ha m_{4}\, \} \ , \cr
&\chi_{k^o}^\pm ~=~ \{\, m_{2}\,, m_{4}\,,m_{3}\,,0\,,m_{26,45}\,;\,
\pm\ha m_{6}\, \} \ , \cr &\chi_{k''}^\pm ~=~ \{\, m_{25,34}\,,
m_{6}\,,m_{5}\,,m_{4}\,,m_{2}\,;\, 0\, \} \ \cr }$$ These multiplets
are depicted on Fig. 5.

\subsec{Doubly reduced types of multiplets}
\subsubsecno=0

\subsubsec{Type R24}

\nt
Here belong multiplets of further reduction, first of reduced type R24 containing 31 ERs/GVMs
which may be obtained formally from the R2 type by
setting ~$m_4 ~=~ 0$. Their signatures are given explicitly by:
\eqna\tablee
$$\eqalignno{
&\chi_b^\pm ~=~ \{\, m_1\,, m_{3}\,,0\,,m_{5}\,,m_6\,;\, \pm\ha
m_\ta\, \} \ , &\tablee{}\cr &\chi_c^\pm ~=~ \{\, m_{1,3}\,,
0\,,0\,,m_{3,5}\,,m_6\,;\, \pm\ha (m_\ta-m_{3})\, \} \ , \cr
&\chi_{c'}^\pm ~=~ \{\, m_{1}\,,m_{35}\,,0\,,0\,,m_{56}\,;\, \pm\ha
(m_\ta-m_{5})\, \} \ , \cr &\chi_{d}^\pm ~=~ \{\,
m_{3}\,,0\,,0\,,m_{1,3,5}\,,m_{6}\,;\, \pm\ha (m_\ta-m_{1,3})\, \} \
, \cr &\chi_{d'}^\pm ~=~ \{\, m_{1}\,,m_{3,56}\,,0\,,0\,,m_{5}\,;\,
\pm\ha (m_\ta-m_{56})\, \} \ , \cr &\chi_{g}^\pm ~=~ \{\,
m_{3}\,,m_{5}\,,0\,,m_{1,3}\,,m_{56}\,;\, \pm\ha m_{3,56}\, \} \ ,
\cr &\chi_{g'}^\pm ~=~ \{\,
m_{1,3}\,,m_{56}\,,0\,,m_{3}\,,m_{5}\,;\,\pm\ha m_{1,3,5}\, \} \ ,
\cr &\chi_{\tg}^\pm ~=~ \{\,
m_{1,3}\,,m_{5}\,,0\,,m_{3}\,,m_{56}\,;\, \pm\ha m_{1,3,56}\, \} \ ,
\cr &\chi_{g^o}^\pm ~=~ \{\,
0\,,m_{5}\,,m_{3}\,,m_{1}\,,m_{3,56}\,;\, \pm\ha m_{56}\, \} \ , \cr
&\chi_{g''}^\pm ~=~ \{\, m_{1,3,5}\,,m_{6}\,,m_{5}\,,m_{3}\,,0\,;\,
\pm\ha m_{1,3}\, \} \ , \cr &\chi_{\hk}^\pm ~=~ \{\,
m_{3}\,,m_{56}\,,0\,,m_{1,3}\,,m_{5}\,;\, \pm\ha m_{3,5}\, \} \ ,
\cr &\chi_{j}^\pm ~=~ \{\, 0\,,m_{56}\,,m_{3}\,,m_{1}\,,m_{3,5}\,;\,
\pm\ha m_{5}\, \} \ , \cr &\chi_{j'}^\pm ~=~ \{\,
m_{3,5}\,,m_{6}\,,m_{5}\,,m_{1,3}\,,0\,;\, \pm\ha m_{3}\, \} \ , \cr
&\chi_{\tk}^\pm ~=~ \{\,
m_{5}\,,m_{6}\,,m_{3,5}\,,m_{1}\,,m_{3}\,;\, 0\, \} \ , \cr
&\chi_{k^o}^\pm ~=~ \{\, 0\,,0\,,m_{3}\,,m_{1}\,,m_{3,56,5}\,;\,
\pm\ha m_{6}\, \} \ , \cr &\chi_{k''}^\pm ~=~ \{\,
m_{1,3,5,3}\,,m_{6}\,,m_{5}\,,0\,,0\,;\, \pm\ha m_{1}\, \} \ \cr }$$
These multiplets are depicted on Fig. 6.

\subsubsec{Type R23}

\nt
Here belong multiplets of another reduction line, first of reduced type R23 containing 35 ERs/GVMs
which may be obtained formally from the R2 type by
setting ~$m_3 ~=~ 0$. Their signatures are given explicitly by:
\eqna\tablf
$$\eqalignno{
&\chi_a^\pm ~=~ \{\, m_1\,, 0\,,m_4\,,m_5\,,m_6\,;\, \pm\ha m_\ta\,
\} \ , &\tablf{}\cr &\chi_c^\pm ~=~ \{\, m_{1}\,,
m_{4}\,,0\,,m_{45}\,,m_6\,;\, \pm\ha (m_\ta-m_{4})\, \} \ , \cr
&\chi_{d}^\pm ~=~ \{\, 0\,,m_{4}\,,0\,,m_{1,45}\,,m_{6}\,;\, \pm\ha
(m_\ta-m_{1,4})\, \} \ , \cr &\chi_{\td}^\pm ~=~ \{\,
m_{1}\,,m_{45}\,,0\,,m_{4}\,,m_{56}\,;\, \pm\ha (m_\ta-m_{45})\, \}
\ , \cr &\chi_{e}^\pm ~=~ \{\,
0\,,m_{45}\,,0\,,m_{1,4}\,,m_{56}\,;\, \pm\ha (m_\ta-m_{1,45})\, \}
\ , \cr &\chi_{e'}^\pm ~=~ \{\,
m_{1}\,,m_{46}\,,0\,,m_{4}\,,m_{5}\,;\, \pm\ha (m_\ta-m_{46})\, \} \
, \cr &\chi_{\te}^\pm ~=~ \{\, 0\,,m_{46}\,,0\,,m_{1,4}\,,m_{5}\,;\,
\pm\ha (m_\ta-m_{1,46})\, \} \ , \cr &\chi_{g}^\pm ~=~ \{\,
m_{4}\,,m_{5}\,,m_{4}\,,m_{1}\,,m_{46}\,;\, \pm\ha m_{46}\, \} \ ,
\cr &\chi_{g'}^\pm ~=~ \{\,
m_{1,4}\,,m_{56}\,,m_{4}\,,0\,,m_{45}\,;\,\pm\ha m_{1,45}\, \} \ ,
\cr &\chi_{\tg}^\pm ~=~ \{\,
m_{1,4}\,,m_{5}\,,m_{4}\,,0\,,m_{46}\,;\, \pm\ha m_{1,46}\, \} \ ,
\cr &\chi_{g''}^\pm ~=~ \{\,
m_{1,45}\,,m_{6}\,,m_{45}\,,0\,,m_{4}\,;\, \pm\ha m_{1,4}\, \} \ ,
\cr &\chi_{\hk}^\pm ~=~ \{\,
m_{4}\,,m_{56}\,,m_{4}\,,m_{1}\,,m_{45}\,;\, \pm\ha m_{45}\, \} \ ,
\cr &\chi_{j^o}^\pm ~=~ \{\, 0\,,m_{45}\,,0\,,m_{1}\,,m_{46,4}\,;\,
\pm\ha m_{56}\, \} \ , \cr &\chi_{k}^\pm ~=~ \{\,
0\,,m_{46}\,,0\,,m_{1}\,,m_{45,4}\,;\, \pm\ha m_{5}\, \} \ , \cr
&\chi_{k'}^\pm ~=~ \{\, m_{45,4}\,,m_{6}\,,m_{5}\,,m_{1,4}\,,0\,;\,
0 \, \} \ , \cr &\chi_{\tk}^\pm ~=~ \{\,
m_{45}\,,m_{6}\,,m_{45}\,,m_{1}\,,m_{4}\,;\, \pm\ha m_{4}\, \} \ ,
\cr &\chi_{k^o}^\pm ~=~ \{\, 0\,,m_{4}\,,0\,,m_{1}\,,m_{46,45}\,;\,
\pm\ha m_{6}\, \} \ , \cr &\chi_{k''}^\pm ~=~ \{\,
m_{1,45,4}\,,m_{6}\,,m_{5}\,,m_{4}\,,0\,;\, \pm\ha m_{1}\, \} \ \cr
}$$ These multiplets are depicted on Fig. 7.

\subsubsec{Type R21}

\nt
Here belong multiplets of another reduction line, first of reduced type R21 containing 35 ERs/GVMs
which may be obtained formally from the R2 type by
setting ~$m_1 ~=~ 0$. Their signatures are given explicitly by:
\eqna\tablg
$$\eqalignno{
&\chi_a^\pm ~=~ \{\, 0\,, m_3\,,m_4\,,m_5\,,m_6\,;\, \pm\ha m_\ta\,
\} \ , &\tablg{}\cr &\chi_b^\pm ~=~ \{\, 0\,,
m_{34}\,,0\,,m_{45}\,,m_6\,;\, \pm\ha (m_\ta-m_{4})\, \} \ , \cr
&\chi_c^\pm ~=~ \{\, m_{3}\,, m_{4}\,,0\,,m_{35}\,,m_6\,;\, \pm\ha
(m_\ta-m_{34})\, \} \ , \cr &\chi_{c'}^\pm ~=~ \{\,
0\,,m_{35}\,,0\,,m_{4}\,,m_{56}\,;\, \pm\ha (m_\ta-m_{45})\, \} \ ,
\cr &\chi_{d'}^\pm ~=~ \{\, 0\,,m_{36}\,,0\,,m_{4}\,,m_{5}\,;\,
\pm\ha (m_\ta-m_{46})\, \} \ , \cr &\chi_{\td}^\pm ~=~ \{\,
m_{3}\,,m_{45}\,,0\,,m_{34}\,,m_{56}\,;\, \pm\ha (m_\ta-m_{35})\, \}
\ , \cr &\chi_{e'}^\pm ~=~ \{\,
m_{3}\,,m_{46}\,,0\,,m_{34}\,,m_{5}\,;\, \pm\ha (m_\ta-m_{36})\, \}
\ , \cr &\chi_{g'}^\pm ~=~ \{\,
m_{34}\,,m_{56}\,,m_{4}\,,m_{3}\,,m_{45}\,;\,\pm\ha m_{35}\, \} \ ,
\cr &\chi_{\tg}^\pm ~=~ \{\,
m_{34}\,,m_{5}\,,m_{4}\,,m_{3}\,,m_{46}\,;\, \pm\ha m_{36}\, \} \ ,
\cr &\chi_{g^o}^\pm ~=~ \{\,
m_{4}\,,m_{5}\,,m_{34}\,,0\,,m_{36}\,;\, \pm\ha m_{46}\, \} \ , \cr
&\chi_{g''}^\pm ~=~ \{\,
m_{35}\,,m_{6}\,,m_{45}\,,m_{3}\,,m_{4}\,;\, \pm\ha m_{34}\, \} \ ,
\cr &\chi_{j}^\pm ~=~ \{\, m_{4}\,,m_{56}\,,m_{34}\,,0\,,m_{35}\,;\,
\pm\ha m_{45}\, \} \ , \cr &\chi_{j^o}^\pm ~=~ \{\,
0\,,m_{45}\,,m_{3}\,,0\,,m_{36,4}\,;\, \pm\ha m_{56}\, \} \ , \cr
&\chi_{k}^\pm ~=~ \{\, 0\,,m_{46}\,,m_{3}\,,0\,,m_{35,4}\,;\, \pm\ha
m_{5}\, \} \ , \cr &\chi_{k'}^\pm ~=~ \{\,
m_{35,4}\,,m_{6}\,,m_{5}\,,m_{34}\,,0\,;\, \pm\ha m_{3}\, \} \ , \cr
&\chi_{\tk}^\pm ~=~ \{\, m_{45}\,,m_{6}\,,m_{35}\,,0\,,m_{34}\,;\,
\pm\ha m_{4}\, \} \ , \cr &\chi_{k^o}^\pm ~=~ \{\,
0\,,m_{4}\,,m_{3}\,,0\,,m_{36,45}\,;\, \pm\ha m_{6}\, \} \ , \cr
&\chi_{k''}^\pm ~=~ \{\, m_{35,34}\,,m_{6}\,,m_{5}\,,m_{4}\,,0\,;\,
0\, \} \ \cr }$$ These multiplets are depicted on Fig. 8.

\subsubsec{Type R41}

\nt
The multiplets of reduced type R41 contain 36 ERs/GVMs and may be obtained formally from the type R1 by
setting ~$m_4 ~=~ 0$ (or from type R4 by
setting ~$m_1 ~=~ 0$). Their signatures are given explicitly by:
\eqna\tablh
$$\eqalignno{
&\chi_0^\pm ~=~ \{\, 0\,, m_3\,,0\,,m_5\,,m_6\,;\, \pm\ha m_\ta\, \}
\ , &\tablh{}\cr &\chi_b^\pm ~=~ \{\, 0\,,
m_{3}\,,m_{2}\,,m_{5}\,,m_6\,;\, \pm\ha (m_\ta-m_{2})\, \} \ , \cr
&\chi_{c'}^\pm ~=~ \{\, 0\,, m_{3,5}\,,m_{2}\,,0\,,m_{56}\,;\,
\pm\ha (m_\ta-m_{2,5})\, \} \ , \cr
 &\chi_{d}^\pm ~=~ \{\, m_{3}\,,
0\,,m_{2}\,,m_{3,5}\,,m_{6}\,;\, \pm\ha (m_\ta-m_{23})\, \} \ , \cr
&\chi_{d'}^\pm ~=~ \{\, 0\,, m_{3,56}\,,m_{2}\,,0\,,m_{5}\,;\,
\pm\ha (m_\ta-m_{2,56})\, \} \ , \cr
 &\chi_{f}^\pm ~=~ \{\, m_{3}\,,
m_{5}\,,m_{2}\,,m_{3}\,,m_{56}\,;\, \pm\ha (m_\ta-m_{23,5})\, \} \ ,
\cr &\chi_{f^o}^\pm ~=~ \{\, 0\,, m_{5}\,,m_{23}\,,0\,,m_{3,56}\,;\,
\pm\ha (m_\ta-m_{23,5,3})\, \} \ , \cr &\chi_{g}^\pm ~=~ \{\,
m_{23}\,, m_{5}\,,0\,,m_{3}\,,m_{2,56}\,;\, \pm\ha m_{3,56}\, \} \ ,
\cr &\chi_{g^o}^\pm ~=~ \{\, m_{2}\,,
m_{5}\,,m_{3}\,,0\,,m_{23,56}\,;\, \pm\ha m_{56}\, \} \ , \cr
&\chi_{h}^\pm ~=~ \{\, 0\,, m_{56}\,,m_{23}\,,0\,,m_{3,5}\,;\,\pm\ha
m_{2,5}\, \} \ , \cr &\chi_{h'}^\pm ~=~ \{\, m_{3,5}\,,
m_{6}\,,m_{2,5}\,,m_{3}\,,0\,;\, \pm\ha m_{23}\, \} \ , \cr
&\chi_{\th}^\pm ~=~ \{\, m_{3}\,,
m_{56}\,,m_{2}\,,m_{3}\,,m_{5}\,;\,\pm\ha m_{23,5}\, \} \ , \cr
&\chi_{\hh}^\pm ~=~ \{\, m_{5}\,, m_{6}\,,m_{23,5}\,,0\,,m_{3}\,;\,
\pm\ha m_{2}\, \} \ , \cr &\chi_{\hk}^\pm ~=~ \{\, m_{23}\,,
m_{56}\,,0\,,m_{3}\,,m_{2,5}\,;\, \pm\ha m_{3,5}\, \} \ , \cr
&\chi_{k}^\pm ~=~ \{\, m_{2}\,, m_{56}\,,m_{3}\,,0\,,m_{23,5}\,;\,
\pm\ha m_{5}\, \} \ , \cr &\chi_{k'}^\pm ~=~ \{\, m_{23,5}\,,
m_{6}\,,m_{5}\,,m_{3}\,,m_{2}\,;\, \pm\ha m_{3}\, \} \ , \cr
&\chi_{\tk}^\pm ~=~ \{\, m_{2,5}\,,
m_{6}\,,m_{3,5}\,,0\,,m_{23}\,;\, 0\, \} \ , \cr &\chi_{k^o}^\pm ~=~
\{\, m_{2}\,, 0\,,m_{3}\,,0\,,m_{23,56,5}\,;\, \pm\ha m_{6}\, \} \ ,
\cr &\chi_{k''}^\pm ~=~ \{\, m_{23,5,3}\,,
m_{6}\,,m_{5}\,,0\,,m_{2}\,;\, 0\, \} \ \cr }$$ These multiplets are
depicted on Fig. 9.

\subsubsec{Type R43}

\nt The multiplets of reduced type R43 contain 31 ERs/GVMs and may
be obtained formally from the R3 type by setting ~$m_4 ~=~ 0$ (or
from type R4 by setting ~$m_3 ~=~ 0$). Their signatures are given
explicitly by: \eqna\tabli
$$\eqalignno{
&\chi_0^\pm ~=~ \{\, m_1\,, 0\,,0\,,m_5\,,m_6\,;\, \pm\ha m_\ta\, \}
\ , &\tabli{}\cr
 &\chi_c^\pm ~=~ \{\, m_{1}\,,
0\,,m_{2}\,,m_{5}\,,m_6\,;\, \pm\ha (m_\ta-m_{2})\, \} \ , \cr
&\chi_{d}^\pm ~=~ \{\, 0\,, 0\,,m_{2}\,,m_{1,5}\,,m_{6}\,;\, \pm\ha
(m_\ta-m_{12})\, \} \ , \cr &\chi_{f}^\pm ~=~ \{\, 0\,,
m_{5}\,,m_{2}\,,m_{1}\,,m_{56}\,;\, \pm\ha (m_\ta-m_{12,5})\, \} \ ,
\cr &\chi_{f'}^\pm ~=~ \{\, m_{1}\,, m_{56}\,,m_{2}\,,0\,,m_{5}\,;\,
\pm\ha (m_\ta-m_{2,56})\, \} \ , \cr &\chi_{\tf}^\pm ~=~ \{\,
m_{1}\,, m_{5}\,,m_{2}\,,0\,,m_{56}\,;\, \pm\ha (m_\ta-m_{2,5})\, \}
\ , \cr &\chi_{f''}^\pm ~=~ \{\, m_{1,5}\,,
m_{6}\,,m_{2,5}\,,0\,,0\,;\, \pm\ha (m_\ta-m_{2,56,5})\, \} \ , \cr
&\chi_{g}^\pm ~=~ \{\, m_{2}\,, m_{5}\,,0\,,m_{1}\,,m_{2,56}\,;\,
\pm\ha m_{56}\, \} \ , \cr &\chi_{g'}^\pm ~=~ \{\, m_{12}\,,
m_{56}\,,0\,,0\,,m_{2,5}\,;\,\pm\ha m_{1,5}\, \} \ , \cr
&\chi_{\tg}^\pm ~=~ \{\, m_{12}\,, m_{5}\,,0\,,0\,,m_{2,56}\,;\,
\pm\ha m_{1,56}\, \} \ , \cr &\chi_{h'}^\pm ~=~ \{\, m_{5}\,,
m_{6}\,,m_{2,5}\,,m_{1}\,,0\,;\, \pm\ha m_{2}\, \} \ , \cr
&\chi_{\th}^\pm ~=~ \{\, 0\,,
m_{56}\,,m_{2}\,,m_{1}\,,m_{5}\,;\,\pm\ha m_{2,5}\, \} \ , \cr
&\chi_{k}^\pm ~=~ \{\, m_{2}\,, m_{56}\,,0\,,m_{1}\,,m_{2,5}\,;\,
\pm\ha m_{5}\, \} \ , \cr &\chi_{\tk}^\pm ~=~ \{\, m_{2,5}\,,
m_{6}\,,m_{5}\,,m_{1}\,,m_{2}\,;\, 0\, \} \ , \cr &\chi_{k^o}^\pm
~=~ \{\, m_{2}\,, 0\,,0\,,m_{1}\,,m_{2,56,5}\,;\, \pm\ha m_{6}\, \}
\ , \cr &\chi_{k''}^\pm ~=~ \{\, m_{12,5}\,,
m_{6}\,,m_{5}\,,0\,,m_{2}\,;\, \pm\ha m_{1}\, \} \ \cr }$$ These
multiplets are depicted on Fig. 10.

\subsubsec{Type R13}

\nt
The multiplets of reduced type R13 contain 29 ERs/GVMs and may be obtained formally from the R1 type by
setting ~$m_3 ~=~ 0$. Their signatures are given explicitly by:
\eqna\tablj
$$\eqalignno{
&\chi_0^\pm ~=~ \{\, 0\,, 0\,,m_4\,,m_5\,,m_6\,;\, \pm\ha m_\ta\, \}
\ , &\tablj{}\cr &\chi_a^\pm ~=~ \{\, 0\,,
0\,,m_{2,4}\,,m_5\,,m_6\,;\, \pm\ha (m_\ta-m_2)\, \} \ ,\cr
 &\chi_{d}^\pm ~=~ \{\, 0\,,
m_{4}\,,m_{2}\,,m_{45}\,,m_{6}\,;\, \pm\ha (m_\ta-m_{2,4})\, \} \ ,
\cr
 &\chi_{e}^\pm ~=~ \{\, 0\,,
m_{45}\,,m_{2}\,,m_{4}\,,m_{56}\,;\, \pm\ha (m_\ta-m_{2,45})\, \} \
, \cr &\chi_{\te}^\pm ~=~ \{\, 0\,,
m_{46}\,,m_{2}\,,m_{4}\,,m_{5}\,;\, \pm\ha (m_\ta-m_{2,46})\, \} \ ,
\cr &\chi_{f}^\pm ~=~ \{\, m_{4}\,,
m_{5}\,,m_{2,4}\,,0\,,m_{46}\,;\, \pm\ha (m_\ta-m_{2,45,4})\, \} \ ,
\cr &\chi_{g}^\pm ~=~ \{\, m_{2,4}\,,
m_{5}\,,m_{4}\,,0\,,m_{2,46}\,;\, \pm\ha m_{46}\, \} \ , \cr
&\chi_{h'}^\pm ~=~ \{\, m_{45}\,, m_{6}\,,m_{2,45}\,,0\,,m_{4}\,;\,
\pm\ha m_{2,4}\, \} \ , \cr &\chi_{\th}^\pm ~=~ \{\, m_{4}\,,
m_{56}\,,m_{2,4}\,,0\,,m_{45}\,;\,\pm\ha m_{2,45}\, \} \ , \cr
&\chi_{\hk}^\pm ~=~ \{\, m_{2,4}\,,
m_{56}\,,m_{4}\,,0\,,m_{2,45}\,;\, \pm\ha m_{45}\, \} \ , \cr
&\chi_{j^o}^\pm ~=~ \{\, m_{2}\,, m_{45}\,,0\,,0\,,m_{2,46,4}\,;\,
\pm\ha m_{56}\, \} \ , \cr &\chi_{k}^\pm ~=~ \{\, m_{2}\,,
m_{46}\,,0\,,0\,,m_{2,45,4}\,;\, \pm\ha m_{5}\, \} \ , \cr
&\chi_{k'}^\pm ~=~ \{\, m_{2,45,4}\,,
m_{6}\,,m_{5}\,,m_{4}\,,m_{2}\,;\, 0\, \} \ , \cr &\chi_{\tk}^\pm
~=~ \{\, m_{2,45}\,, m_{6}\,,m_{45}\,,0\,,m_{2,4}\,;\, \pm\ha
m_{4}\, \} \ , \cr &\chi_{k^o}^\pm ~=~ \{\, m_{2}\,,
m_{4}\,,0\,,0\,,m_{2,46,45}\,;\, \pm\ha m_{6}\, \} \ \cr
 }$$
These multiplets are depicted on Fig. 11.

\subsubsec{Type R15}

\nt
The multiplets of reduced type R15 contain 34 ERs/GVMs and may be obtained formally from the R1 type by
setting ~$m_5 ~=~ 0$. Their signatures are given explicitly by:
\eqna\tablk
$$\eqalignno{
&\chi_0^\pm ~=~ \{\, 0\,, m_3\,,m_4\,,0\,,m_6\,;\, \pm\ha m_\ta\, \}
\ , &\tablk{}\cr &\chi_a^\pm ~=~ \{\, 0\,,
m_3\,,m_{2,4}\,,0\,,m_6\,;\, \pm\ha (m_\ta-m_2)\, \} \ ,\cr
&\chi_{c'}^\pm ~=~ \{\, 0\,, m_{34}\,,m_{2}\,,m_{4}\,,m_{6}\,;\,
\pm\ha (m_\ta-m_{2,4})\, \} \ , \cr &\chi_{d'}^\pm ~=~ \{\, 0\,,
m_{34,6}\,,m_{2}\,,m_{4}\,,0\,;\, \pm\ha (m_\ta-m_{2,4,6})\, \} \ ,
\cr
 &\chi_{e}^\pm ~=~ \{\, m_{3}\,,
m_{4}\,,m_{2}\,,m_{34}\,,m_{6}\,;\, \pm\ha (m_\ta-m_{24})\, \} \ ,
\cr &\chi_{\te}^\pm ~=~ \{\, m_{3}\,,
m_{4,6}\,,m_{2}\,,m_{34}\,,0\,;\, \pm\ha (m_\ta-m_{24,6})\, \} \ ,
\cr &\chi_{f}^\pm ~=~ \{\, m_{34}\,,
0\,,m_{2,4}\,,m_{3}\,,m_{4,6}\,;\, \pm\ha (m_\ta-m_{24,4})\, \} \ ,
\cr &\chi_{f^o}^\pm ~=~ \{\, m_{4}\,, 0\,,m_{24}\,,0\,,m_{34,6}\,;\,
\pm\ha (m_\ta-m_{24,34})\, \} \ , \cr &\chi_{g}^\pm ~=~ \{\,
m_{24}\,, 0\,,m_{4}\,,m_{3}\,,m_{2,4,6}\,;\, \pm\ha m_{34,6}\, \} \
, \cr &\chi_{g^o}^\pm ~=~ \{\, m_{2,4}\,,
0\,,m_{34}\,,0\,,m_{24,6}\,;\, \pm\ha m_{4,6}\, \} \ , \cr
&\chi_{h}^\pm ~=~ \{\, m_{4}\,,
m_{6}\,,m_{24}\,,0\,,m_{34}\,;\,\pm\ha m_{2,4}\, \} \ , \cr
&\chi_{\th}^\pm ~=~ \{\, m_{34}\,,
m_{6}\,,m_{2,4}\,,m_{3}\,,m_{4}\,;\,\pm\ha m_{24}\, \} \ , \cr
&\chi_{\hk}^\pm ~=~ \{\, m_{24}\,,
m_{6}\,,m_{4}\,,m_{3}\,,m_{2,4}\,;\, \pm\ha m_{34}\, \} \ , \cr
&\chi_{k}^\pm ~=~ \{\, m_{2}\,, m_{4,6}\,,m_{3}\,,0\,,m_{24,4}\,;\,
0\, \} \ , \cr &\chi_{k'}^\pm ~=~ \{\, m_{24,4}\,,
m_{6}\,,0\,,m_{34}\,,m_{2}\,;\, \pm\ha m_{3}\, \} \ , \cr
&\chi_{\tk}^\pm ~=~ \{\, m_{2,4}\,, m_{6}\,,m_{34}\,,0\,,m_{24}\,;\,
\pm\ha m_{4}\, \} \ , \cr &\chi_{k^o}^\pm ~=~ \{\, m_{2}\,,
m_{4}\,,m_{3}\,,0\,,m_{24,6,4}\,;\, \pm\ha m_{6}\, \} \ , \cr
&\chi_{k''}^\pm ~=~ \{\, m_{24,34}\,, m_{6}\,,0\,,m_{4}\,,m_{2}\,;\,
0\, \} \ \cr }$$ These multiplets are depicted on Fig. 12.

\subsubsec{Type R16}

\nt The multiplets of reduced type R16 contain 34 ERs/GVMs and may
be obtained formally from the R1 type by setting ~$m_6 ~=~ 0$. Their
signatures are given explicitly by: \eqna\tabll
$$\eqalignno{
&\chi_0^\pm ~=~ \{\, 0\,, m_3\,,m_4\,,m_5\,,0\,;\, \pm\ha m_\ta\, \}
\ , &\tabll{}\cr &\chi_a^\pm ~=~ \{\, 0\,,
m_3\,,m_{2,4}\,,m_5\,,0\,;\, \pm\ha (m_\ta-m_2)\, \} \ ,\cr
&\chi_b^\pm ~=~ \{\, 0\,, m_{34}\,,m_{2}\,,m_{45}\,,0\,;\, \pm\ha
(m_\ta-m_{2,4})\, \} \ , \cr
 &\chi_{d}^\pm ~=~ \{\, m_{3}\,,
m_{4}\,,m_{2}\,,m_{35}\,,0\,;\, \pm\ha (m_\ta-m_{24})\, \} \ , \cr
&\chi_{d'}^\pm ~=~ \{\, 0\,, m_{35}\,,m_{2}\,,m_{4}\,,m_{5}\,;\,
\pm\ha (m_\ta-m_{2,45})\, \} \ , \cr
 &\chi_{\te}^\pm ~=~ \{\, m_{3}\,,
m_{45}\,,m_{2}\,,m_{34}\,,m_{5}\,;\, \pm\ha (m_\ta-m_{25})\, \} \ ,
\cr &\chi_{h}^\pm ~=~ \{\, m_{4}\,,
m_{5}\,,m_{24}\,,0\,,m_{35}\,;\,\pm\ha m_{2,45}\, \} \ , \cr
&\chi_{h'}^\pm ~=~ \{\, m_{35}\,, 0\,,m_{2,45}\,,m_{3}\,,m_{4}\,;\,
\pm\ha m_{24}\, \} \ , \cr &\chi_{\th}^\pm ~=~ \{\, m_{34}\,,
m_{5}\,,m_{2,4}\,,m_{3}\,,m_{45}\,;\,\pm\ha m_{25}\, \} \ , \cr
&\chi_{\hh}^\pm ~=~ \{\, m_{45}\,, 0\,,m_{25}\,,0\,,m_{34}\,;\,
\pm\ha m_{2,4}\, \} \ , \cr &\chi_{\hk}^\pm ~=~ \{\, m_{24}\,,
m_{5}\,,m_{4}\,,m_{3}\,,m_{2,45}\,;\, \pm\ha m_{35}\, \} \ , \cr
&\chi_{j}^\pm ~=~ \{\, m_{2,4}\,, m_{5}\,,m_{34}\,,0\,,m_{25}\,;\,
\pm\ha m_{45}\, \} \ , \cr &\chi_{j'}^\pm ~=~ \{\, m_{25}\,,
0\,,m_{45}\,,m_{3}\,,m_{2,4}\,;\, \pm\ha m_{34}\, \} \ , \cr
&\chi_{k}^\pm ~=~ \{\, m_{2}\,, m_{45}\,,m_{3}\,,0\,,m_{25,4}\,;\,
\pm\ha m_{5}\, \} \ , \cr &\chi_{k'}^\pm ~=~ \{\, m_{25,4}\,,
0\,,m_{5}\,,m_{34}\,,m_{2}\,;\, \pm\ha m_{3}\, \} \ , \cr
&\chi_{\tk}^\pm ~=~ \{\, m_{2,45}\,, 0\,,m_{35}\,,0\,,m_{24}\,;\,
\pm\ha m_{4}\, \} \ , \cr &\chi_{k^o}^\pm ~=~ \{\, m_{2}\,,
m_{4}\,,m_{3}\,,0\,,m_{25,45}\,;\, 0\, \} \ , \cr &\chi_{k''}^\pm
~=~ \{\, m_{25,34}\,, 0\,,m_{5}\,,m_{4}\,,m_{2}\,;\, 0\, \} \ \cr
}$$ These multiplets are depicted on Fig. 13.

\subsubsec{Type R35}

\nt
The multiplets of reduced type R35 contain 34 ERs/GVMs and may be obtained formally from the R3 type by
setting ~$m_5 ~=~ 0$. Their signatures are given explicitly by:
\eqna\tablm
$$\eqalignno{
&\chi_0^\pm ~=~ \{\, m_1\,, 0\,,m_4\,,0\,,m_6\,;\, \pm\ha m_\ta\, \}
\ , &\tablm{}\cr &\chi_a^\pm ~=~ \{\, m_1\,,
0\,,m_{2,4}\,,0\,,m_6\,;\, \pm\ha (m_\ta-m_2)\, \} \ , \cr
 &\chi_{\td}^\pm ~=~ \{\, m_{1}\,,
m_{4}\,,m_{2}\,,m_{4}\,,m_{6}\,;\, \pm\ha (m_\ta-m_{2,4})\, \} \ ,
\cr &\chi_{e}^\pm ~=~ \{\, 0\,, m_{4}\,,m_{2}\,,m_{1,4}\,,m_{6}\,;\,
\pm\ha (m_\ta-m_{12,4})\, \} \ , \cr &\chi_{e'}^\pm ~=~ \{\,
m_{1}\,, m_{4,6}\,,m_{2}\,,m_{4}\,,0\,;\, \pm\ha (m_\ta-m_{2,4,6})\,
\} \ , \cr &\chi_{\te}^\pm ~=~ \{\, 0\,,
m_{4,6}\,,m_{2}\,,m_{1,4}\,,0\,;\, \pm\ha (m_\ta-m_{12,4,6})\, \} \
, \cr &\chi_{f}^\pm ~=~ \{\, m_{4}\,,
0\,,m_{2,4}\,,m_{1}\,,m_{4,6}\,;\, \pm\ha (m_\ta-m_{12,4,4})\, \} \
, \cr &\chi_{f'}^\pm ~=~ \{\, m_{1,4}\,,
m_{6}\,,m_{2,4}\,,0\,,m_{4}\,;\, \pm\ha (m_\ta-m_{2,4,6,4})\, \} \ ,
\cr &\chi_{\tf}^\pm ~=~ \{\, m_{1,4}\,,
0\,,m_{2,4}\,,0\,,m_{4,6}\,;\, \pm\ha (m_\ta-m_{2,4,4})\, \} \ , \cr
&\chi_{g}^\pm ~=~ \{\, m_{2,4}\,, 0\,,m_{4}\,,m_{1}\,,m_{2,4,6}\,;\,
\pm\ha m_{4,6}\, \} \ , \cr &\chi_{g'}^\pm ~=~ \{\, m_{12,4}\,,
m_{6}\,,m_{4}\,,0\,,m_{2,4}\,;\,\pm\ha m_{1,4}\, \} \ , \cr
&\chi_{\tg}^\pm ~=~ \{\, m_{12,4}\,, 0\,,m_{4}\,,0\,,m_{2,4,6}\,;\,
\pm\ha m_{1,4,6}\, \} \ , \cr &\chi_{\th}^\pm ~=~ \{\, m_{4}\,,
m_{6}\,,m_{2,4}\,,m_{1}\,,m_{45}\,;\,\pm\ha m_{2,4}\, \} \ , \cr
&\chi_{k}^\pm ~=~ \{\, m_{2}\,, m_{4,6}\,,0\,,m_{1}\,,m_{2,4,4}\,;\,
0\, \} \ , \cr &\chi_{k'}^\pm ~=~ \{\, m_{2,4,4}\,,
m_{6}\,,0\,,m_{1,4}\,,m_{2}\,;\, 0\, \} \ , \cr &\chi_{\tk}^\pm ~=~
\{\, m_{2,4}\,, m_{6}\,,m_{4}\,,m_{1}\,,m_{2,4}\,;\, \pm\ha m_{4}\,
\} \ , \cr &\chi_{k^o}^\pm ~=~ \{\, m_{2}\,,
m_{4}\,,0\,,m_{1}\,,m_{2,4,6,4}\,;\, \pm\ha m_{6}\, \} \ , \cr
&\chi_{k''}^\pm ~=~ \{\, m_{12,4,4}\,,
m_{6}\,,0\,,m_{4}\,,m_{2}\,;\, \pm\ha m_{1}\, \} \ \cr }$$ These
multiplets are depicted on Fig. 14.

\subsec{Triply reduced type of multiplets}
\subsubsecno=0

\nt
Here we give only multiplets that contain at least one non-limit ER,
since the others would not be relevant for the applications.

\subsubsec{Type R235}

\nt
Next comes reduced type R235 containing 24 ERs/GVMs
which may be obtained formally from the R23 type by
setting ~$m_5 ~=~ 0$. Their signatures are given explicitly by:
\eqna\tablan
$$\eqalignno{
&\chi_a^\pm ~=~ \{\, m_1\,, 0\,,m_4\,,0\,,m_6\,;\, \pm\ha m_\ta\, \}
\ , &\tablan{}\cr &\chi_{\td}^\pm ~=~ \{\,
m_{1}\,,m_{4}\,,0\,,m_{4}\,,m_{6}\,;\, \pm\ha (m_\ta-m_{4})\, \} \ ,
\cr &\chi_{e}^\pm ~=~ \{\, 0\,,m_{4}\,,0\,,m_{1,4}\,,m_{6}\,;\,
\pm\ha (m_\ta-m_{1,4})\, \} \ , \cr &\chi_{e'}^\pm ~=~ \{\,
m_{1}\,,m_{4,6}\,,0\,,m_{4}\,,0\,;\, \pm\ha (m_\ta-m_{4,6})\, \} \ ,
\cr &\chi_{\te}^\pm ~=~ \{\, 0\,,m_{4,6}\,,0\,,m_{1,4}\,,0\,;\,
\pm\ha (m_\ta-m_{1,4,6})\, \} \ , \cr &\chi_{g}^\pm ~=~ \{\,
m_{4}\,,0\,,m_{4}\,,m_{1}\,,m_{4,6}\,;\, \pm\ha m_{4,6}\, \} \ , \cr
&\chi_{g'}^\pm ~=~ \{\,
m_{1,4}\,,m_{6}\,,m_{4}\,,0\,,m_{4}\,;\,\pm\ha m_{1,4}\, \} \ , \cr
&\chi_{\tg}^\pm ~=~ \{\, m_{1,4}\,,0\,,m_{4}\,,0\,,m_{4,6}\,;\,
\pm\ha m_{1,4,6}\, \} \ , \cr &\chi_{k}^\pm ~=~ \{\,
0\,,m_{4,6}\,,0\,,m_{1}\,,m_{4,4}\,;\, 0\, \} \ , \cr &\chi_{k'}^\pm
~=~ \{\, m_{4,4}\,,m_{6}\,,0\,,m_{1,4}\,,0\,;\, 0 \, \} \ , \cr
&\chi_{\tk}^\pm ~=~ \{\, m_{4}\,,m_{6}\,,m_{4}\,,m_{1}\,,m_{4}\,;\,
\pm\ha m_{4}\, \} \ , \cr &\chi_{k^o}^\pm ~=~ \{\,
0\,,m_{4}\,,0\,,m_{1}\,,m_{4,4,6}\,;\, \pm\ha m_{6}\, \} \ , \cr
&\chi_{k''}^\pm ~=~ \{\, m_{1,4,4}\,,m_{6}\,,0\,,m_{4}\,,0\,;\,
\pm\ha m_{1}\, \} \ \cr }$$ These multiplets are depicted on Fig.
15.

\subsubsec{Type R215}

\nt
Next comes reduced type R215 containing 24 ERs/GVMs
which may be obtained formally from the R21 type by
setting ~$m_5 ~=~ 0$. Their signatures are given explicitly by:
\eqna\tablo
$$\eqalignno{
&\chi_a^\pm ~=~ \{\, 0\,, m_3\,,m_4\,,0\,,m_6\,;\, \pm\ha m_\ta\, \}
\ , &\tablo{}\cr &\chi_{c'}^\pm ~=~ \{\,
0\,,m_{34}\,,0\,,m_{4}\,,m_{6}\,;\, \pm\ha (m_\ta-m_{4})\, \} \ ,
\cr &\chi_{d'}^\pm ~=~ \{\, 0\,,m_{34,6}\,,0\,,m_{4}\,,0\,;\, \pm\ha
(m_\ta-m_{4,6})\, \} \ , \cr &\chi_{\td}^\pm ~=~ \{\,
m_{3}\,,m_{4}\,,0\,,m_{34}\,,m_{6}\,;\, \pm\ha (m_\ta-m_{34})\, \} \
, \cr &\chi_{e'}^\pm ~=~ \{\, m_{3}\,,m_{4,6}\,,0\,,m_{34}\,,0\,;\,
\pm\ha (m_\ta-m_{34,6})\, \} \ , \cr &\chi_{g'}^\pm ~=~ \{\,
m_{34}\,,m_{6}\,,m_{4}\,,m_{3}\,,m_{4}\,;\,\pm\ha m_{34}\, \} \ ,
\cr &\chi_{\tg}^\pm ~=~ \{\,
m_{34}\,,0\,,m_{4}\,,m_{3}\,,m_{4,6}\,;\, \pm\ha m_{34,6}\, \} \ ,
\cr &\chi_{g^o}^\pm ~=~ \{\, m_{4}\,,0\,,m_{34}\,,0\,,m_{34,6}\,;\,
\pm\ha m_{4,6}\, \} \ , \cr &\chi_{k}^\pm ~=~ \{\,
0\,,m_{4,6}\,,m_{3}\,,0\,,m_{34,4}\,;\, 0\, \} \ , \cr
&\chi_{k'}^\pm ~=~ \{\, m_{34,4}\,,m_{6}\,,0\,,m_{34}\,,0\,;\,
\pm\ha m_{3}\, \} \ , \cr &\chi_{\tk}^\pm ~=~ \{\,
m_{4}\,,m_{6}\,,m_{34}\,,0\,,m_{34}\,;\, \pm\ha m_{4}\, \} \ , \cr
&\chi_{k^o}^\pm ~=~ \{\, 0\,,m_{4}\,,m_{3}\,,0\,,m_{34,6,4}\,;\,
\pm\ha m_{6}\, \} \ , \cr &\chi_{k''}^\pm ~=~ \{\,
m_{34,34}\,,m_{6}\,,0\,,m_{4}\,,0\,;\, 0\, \} \ \cr }$$ These
multiplets are depicted on Fig. 16.

\subsubsec{Type R216}

\nt
Next comes reduced type R216 containing 26 ERs/GVMs
which may be obtained formally from the R21 type by
setting ~$m_6 ~=~ 0$. Their signatures are given explicitly by:
\eqna\tablp
$$\eqalignno{
&\chi_a^\pm ~=~ \{\, 0\,, m_3\,,m_4\,,m_5\,,0\,;\, \pm\ha m_\ta\, \}
\ , &\tablp{}\cr &\chi_b^\pm ~=~ \{\, 0\,,
m_{34}\,,0\,,m_{45}\,,0\,;\, \pm\ha (m_\ta-m_{4})\, \} \ , \cr
&\chi_c^\pm ~=~ \{\, m_{3}\,, m_{4}\,,0\,,m_{35}\,,0\,;\, \pm\ha
(m_\ta-m_{34})\, \} \ , \cr &\chi_{c'}^\pm ~=~ \{\,
0\,,m_{35}\,,0\,,m_{4}\,,m_{5}\,;\, \pm\ha (m_\ta-m_{45})\, \} \ ,
\cr &\chi_{d'}^\pm ~=~ \{\, 0\,,m_{35}\,,0\,,m_{4}\,,m_{5}\,;\,
\pm\ha (m_\ta-m_{45})\, \} \ , \cr &\chi_{e'}^\pm ~=~ \{\,
m_{3}\,,m_{45}\,,0\,,m_{34}\,,m_{5}\,;\, \pm\ha (m_\ta-m_{35})\, \}
\ , \cr &\chi_{g'}^\pm ~=~ \{\,
m_{34}\,,m_{5}\,,m_{4}\,,m_{3}\,,m_{45}\,;\,\pm\ha m_{35}\, \} \ ,
\cr &\chi_{g''}^\pm ~=~ \{\,
m_{35}\,,0\,,m_{45}\,,m_{3}\,,m_{4}\,;\, \pm\ha m_{34}\, \} \ , \cr
&\chi_{j}^\pm ~=~ \{\, m_{4}\,,m_{5}\,,m_{34}\,,0\,,m_{35}\,;\,
\pm\ha m_{45}\, \} \ , \cr &\chi_{k}^\pm ~=~ \{\,
0\,,m_{45}\,,m_{3}\,,0\,,m_{35,4}\,;\, \pm\ha m_{5}\, \} \ , \cr
&\chi_{k'}^\pm ~=~ \{\, m_{35,4}\,,0\,,m_{5}\,,m_{34}\,,0\,;\,
\pm\ha m_{3}\, \} \ , \cr &\chi_{\tk}^\pm ~=~ \{\,
m_{45}\,,0\,,m_{35}\,,0\,,m_{34}\,;\, \pm\ha m_{4}\, \} \ , \cr
&\chi_{k^o}^\pm ~=~ \{\, 0\,,m_{4}\,,m_{3}\,,0\,,m_{35,45}\,;\, 0\,
\} \ , \cr &\chi_{k''}^\pm ~=~ \{\,
m_{35,34}\,,0\,,m_{5}\,,m_{4}\,,0\,;\, 0\, \} \ \cr }$$ These
multiplets are depicted on Fig. 17.

\subsubsec{Type R416}

\nt
The multiplets of reduced type R416 contain 25 ERs/GVMs and may be obtained formally from the type R41 by
setting ~$m_6 ~=~ 0$. Their signatures are given explicitly by:
\eqna\tablq
$$\eqalignno{
&\chi_0^\pm ~=~ \{\, 0\,, m_3\,,0\,,m_5\,,0\,;\, \pm\ha m_\ta\, \} \
, &\tablq{}\cr &\chi_b^\pm ~=~ \{\, 0\,,
m_{3}\,,m_{2}\,,m_{5}\,,0\,;\, \pm\ha (m_\ta-m_{2})\, \} \ , \cr
&\chi_{c'}^\pm ~=~ \{\, 0\,, m_{3,5}\,,m_{2}\,,0\,,m_{5}\,;\, \pm\ha
(m_\ta-m_{2,5})\, \} \ , \cr
 &\chi_{d}^\pm ~=~ \{\, m_{3}\,,
0\,,m_{2}\,,m_{3,5}\,,0\,;\, \pm\ha (m_\ta-m_{23})\, \} \ , \cr
 &\chi_{f}^\pm ~=~ \{\, m_{3}\,,
m_{5}\,,m_{2}\,,m_{3}\,,m_{5}\,;\, \pm\ha m_{23,5} \, \} \ , \cr
&\chi_{f^o}^\pm ~=~ \{\, 0\,, m_{5}\,,m_{23}\,,0\,,m_{3,5}\,;\,
\pm\ha m_{2,5} \, \} \ , \cr &\chi_{g}^\pm ~=~ \{\, m_{23}\,,
m_{5}\,,0\,,m_{3}\,,m_{2,5}\,;\, \pm\ha m_{3,5}\, \} \ , \cr
&\chi_{h'}^\pm ~=~ \{\, m_{3,5}\,, 0\,,m_{2,5}\,,m_{3}\,,0\,;\,
\pm\ha m_{23}\, \} \ , \cr &\chi_{\hh}^\pm ~=~ \{\, m_{5}\,,
0\,,m_{23,5}\,,0\,,m_{3}\,;\, \pm\ha m_{2}\, \} \ , \cr
&\chi_{k}^\pm ~=~ \{\, m_{2}\,, m_{5}\,,m_{3}\,,0\,,m_{23,5}\,;\,
\pm\ha m_{5}\, \} \ , \cr &\chi_{k'}^\pm ~=~ \{\, m_{23,5}\,,
0\,,m_{5}\,,m_{3}\,,m_{2}\,;\, \pm\ha m_{3}\, \} \ , \cr
&\chi_{\tk}^\pm ~=~ \{\, m_{2,5}\,, 0\,,m_{3,5}\,,0\,,m_{23}\,;\,
0\, \} \ , \cr &\chi_{k^o}^\pm ~=~ \{\, m_{2}\,,
0\,,m_{3}\,,0\,,m_{23,5,5}\,;\, 0\, \} \ , \cr &\chi_{k''}^\pm ~=~
\{\, m_{23,5,3}\,, 0\,,m_{5}\,,0\,,m_{2}\,;\, 0\, \} \ \cr }$$ These
multiplets are depicted on Fig. 18.

%\np

\newsec{Outlook}

\nt In the present paper we continued the programme outlined in
\Dobinv\ on the example of the non-compact group $E_{6(-14)}\,$.
Similar explicit descriptions are planned for the other non-compact
groups, in particular those with highest/lowest weight
representations. We plan also to extend these considerations to the
supersymmetric cases and also to the quantum group setting. Such
considerations are expected to be very useful for applications to
string theory and integrable models, cf., e.g., \Witten.

\bigskip

\nt {\bf Acknowledgements.}

\nt The author would like to thank for hospitality the Abdus Salam
International Centre for Theoretical Physics, where part of the work
was done. The author was  supported in part by
  the European RTN network {\it ``Forces-Universe''} (contract
No.{\it MRTN-CT-2004-005104}), by Bulgarian NSF grant  {\it DO
02-257}, and by the Alexander von Humboldt Foundation in the
framework of the Clausthal-Leipzig-Sofia Cooperation.

\parskip=0pt
\listrefs
\np

\fig{}{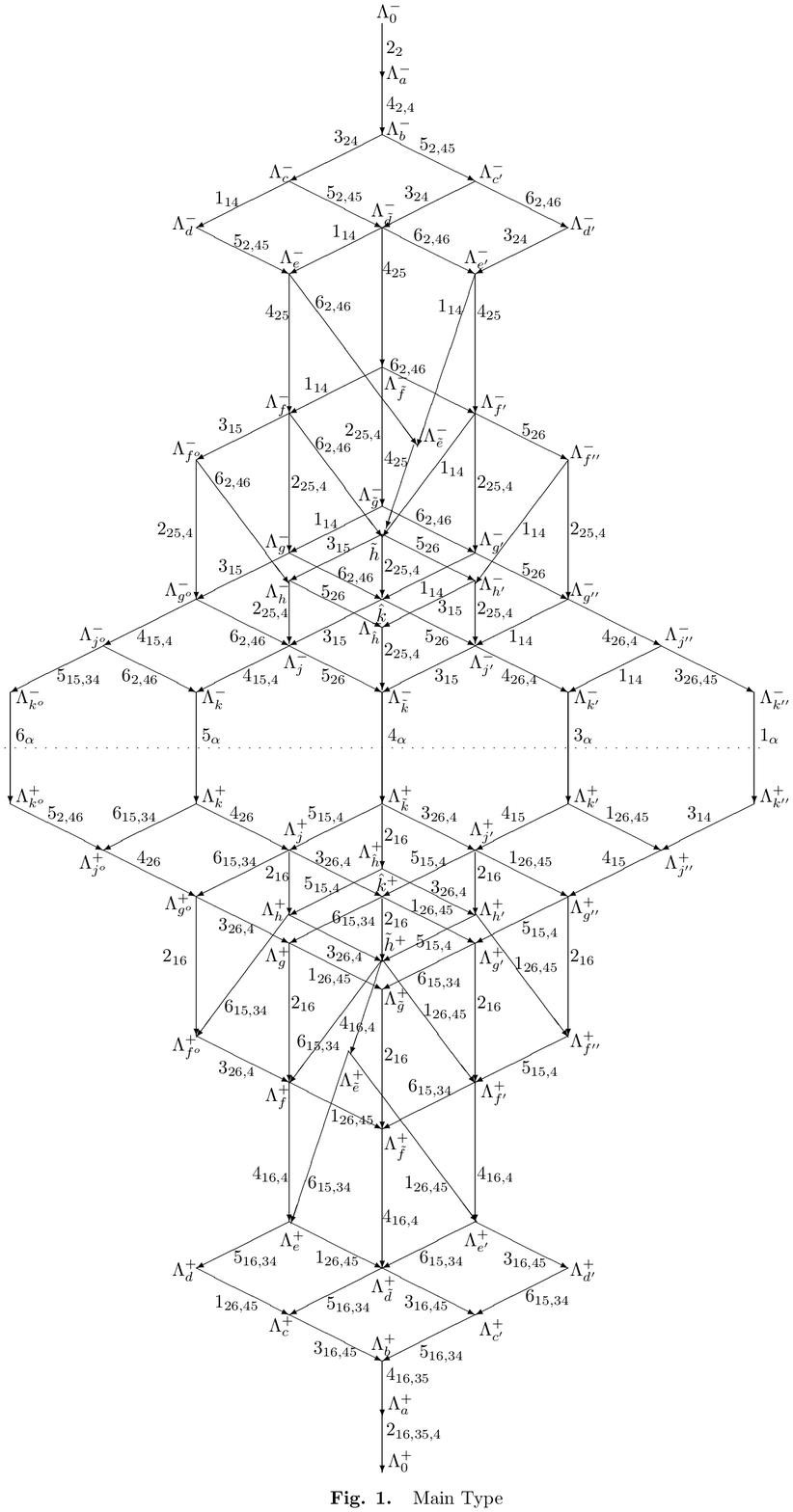}{10cm}

\np

\fig{}{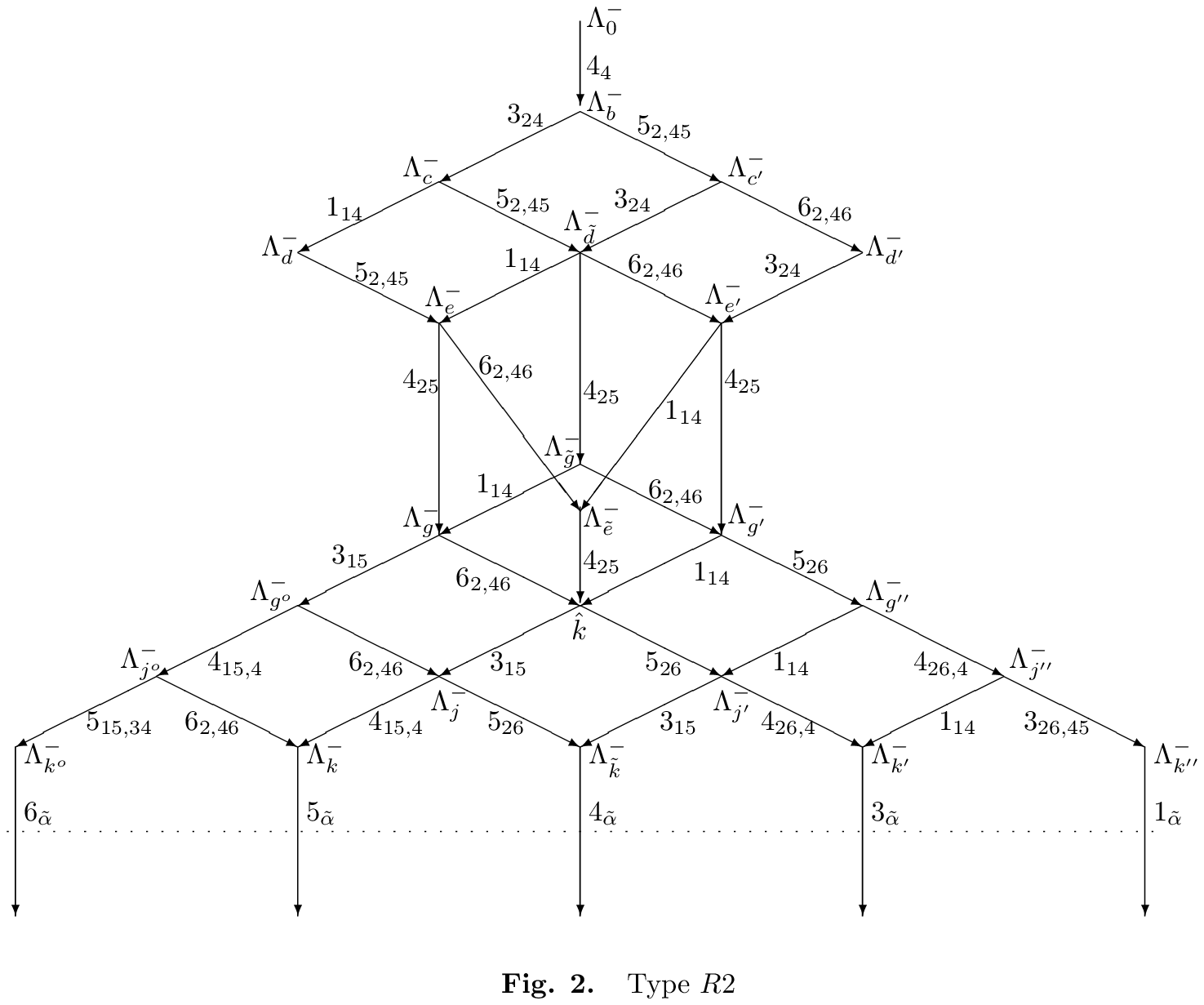}{6.7cm}
\fig{}{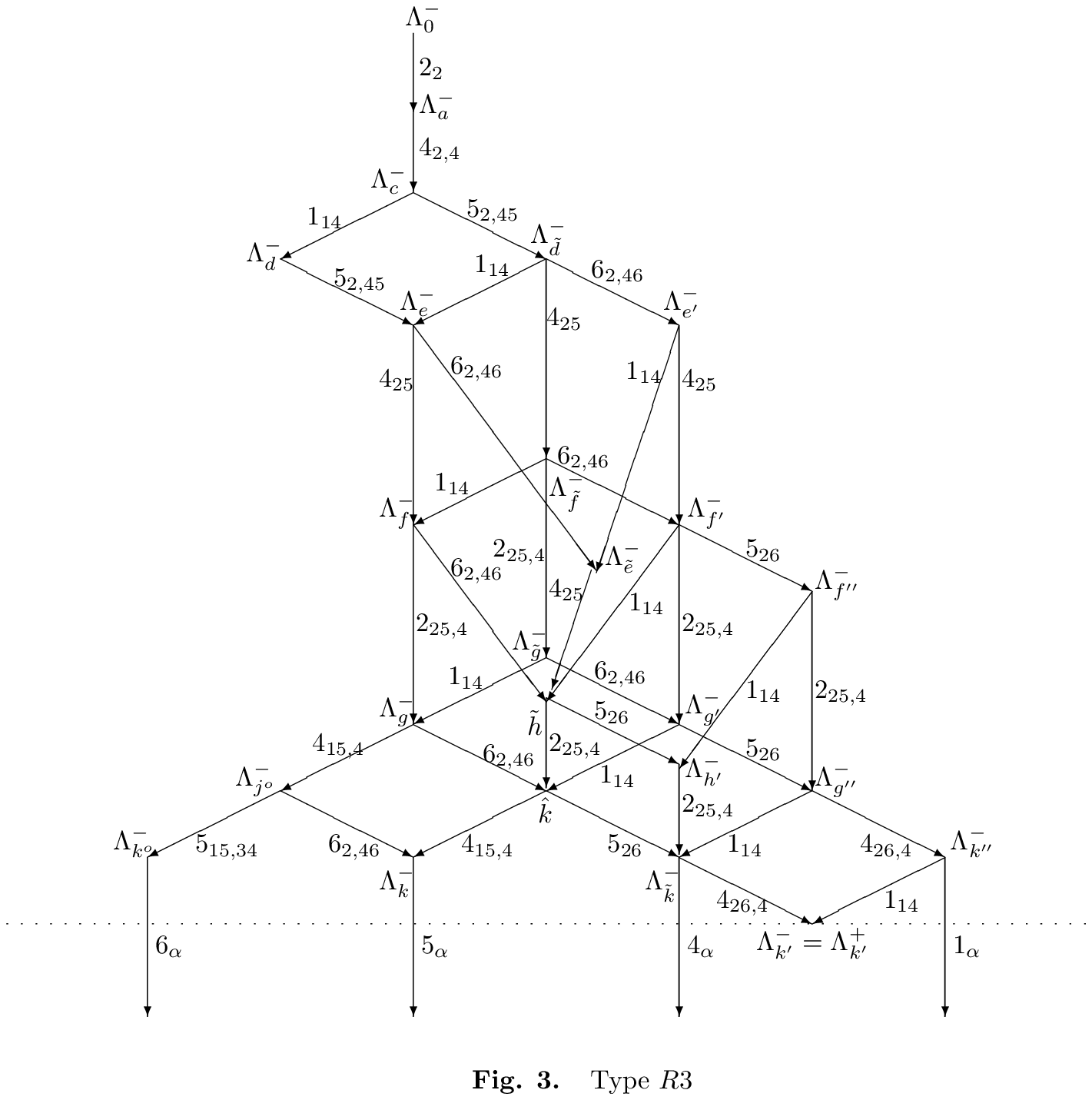}{5.4cm}
\fig{}{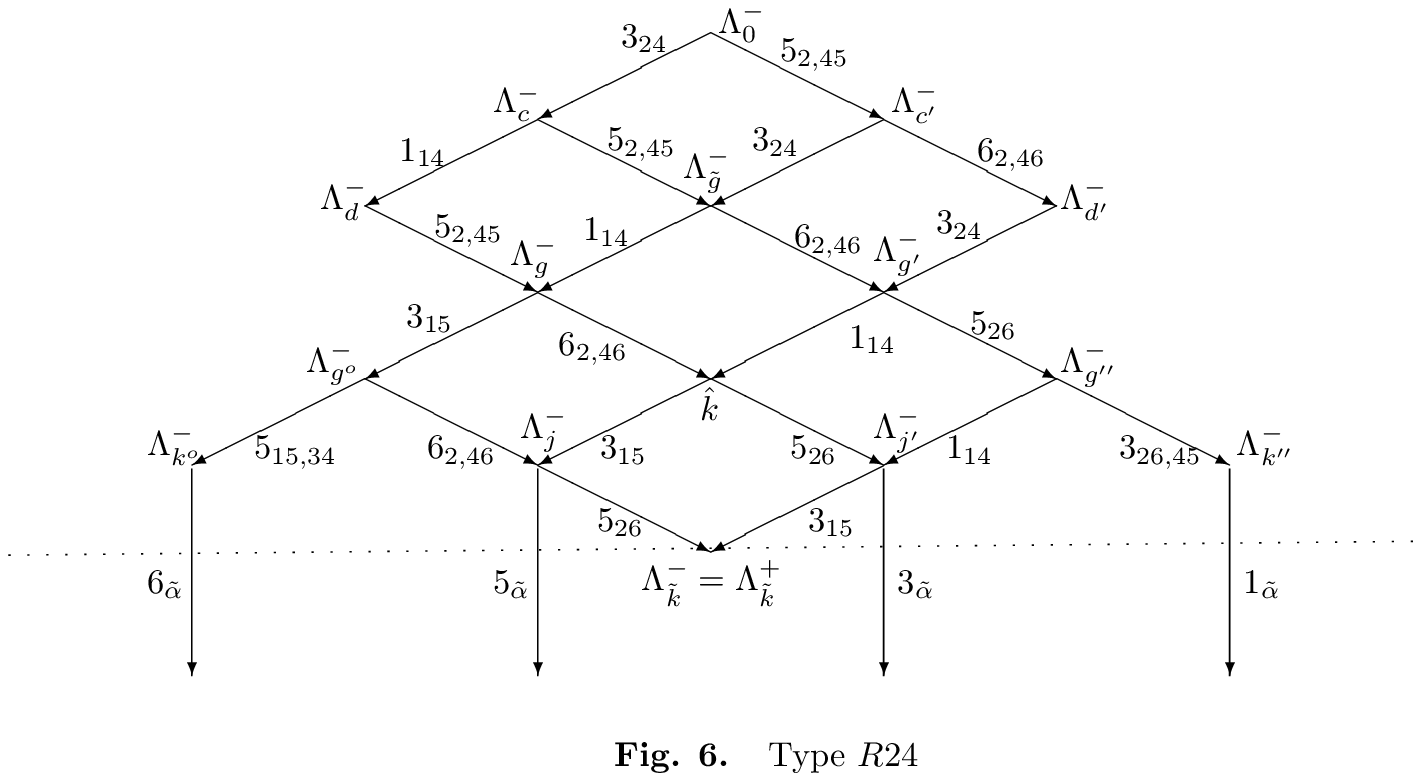}{5.4cm}

\np

\fig{}{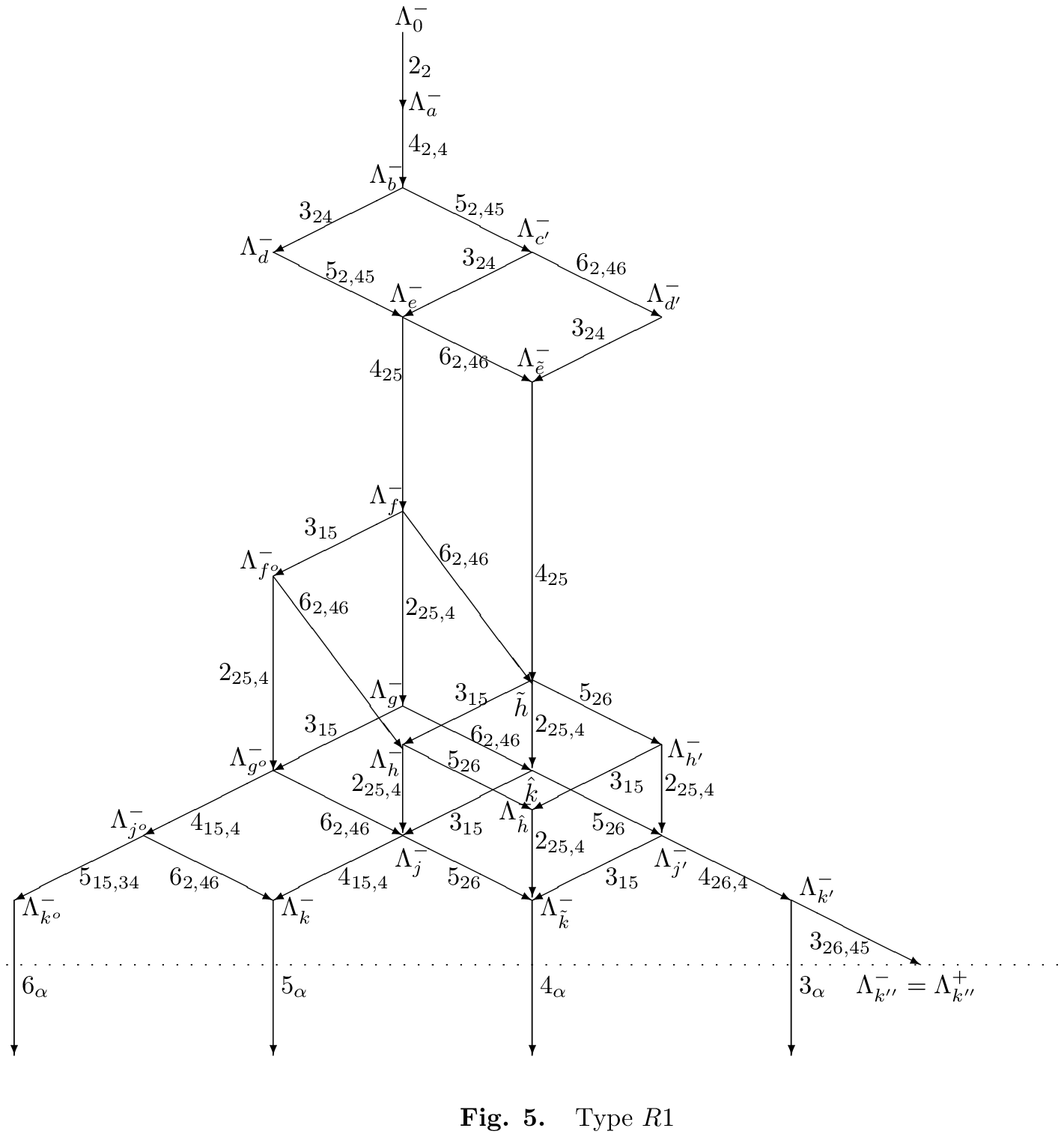}{6.7cm}
\fig{}{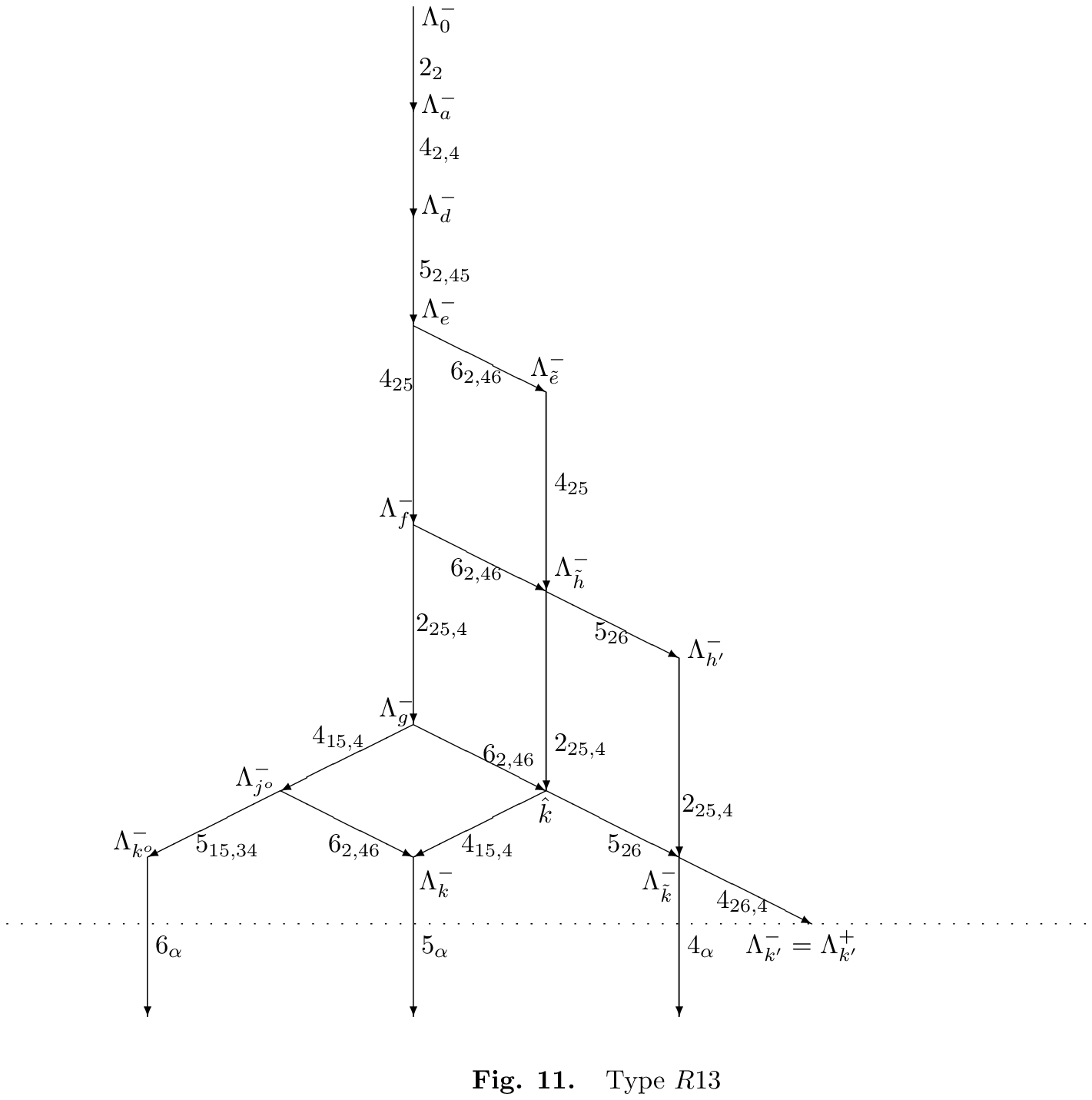}{4.5cm}

\np

\fig{}{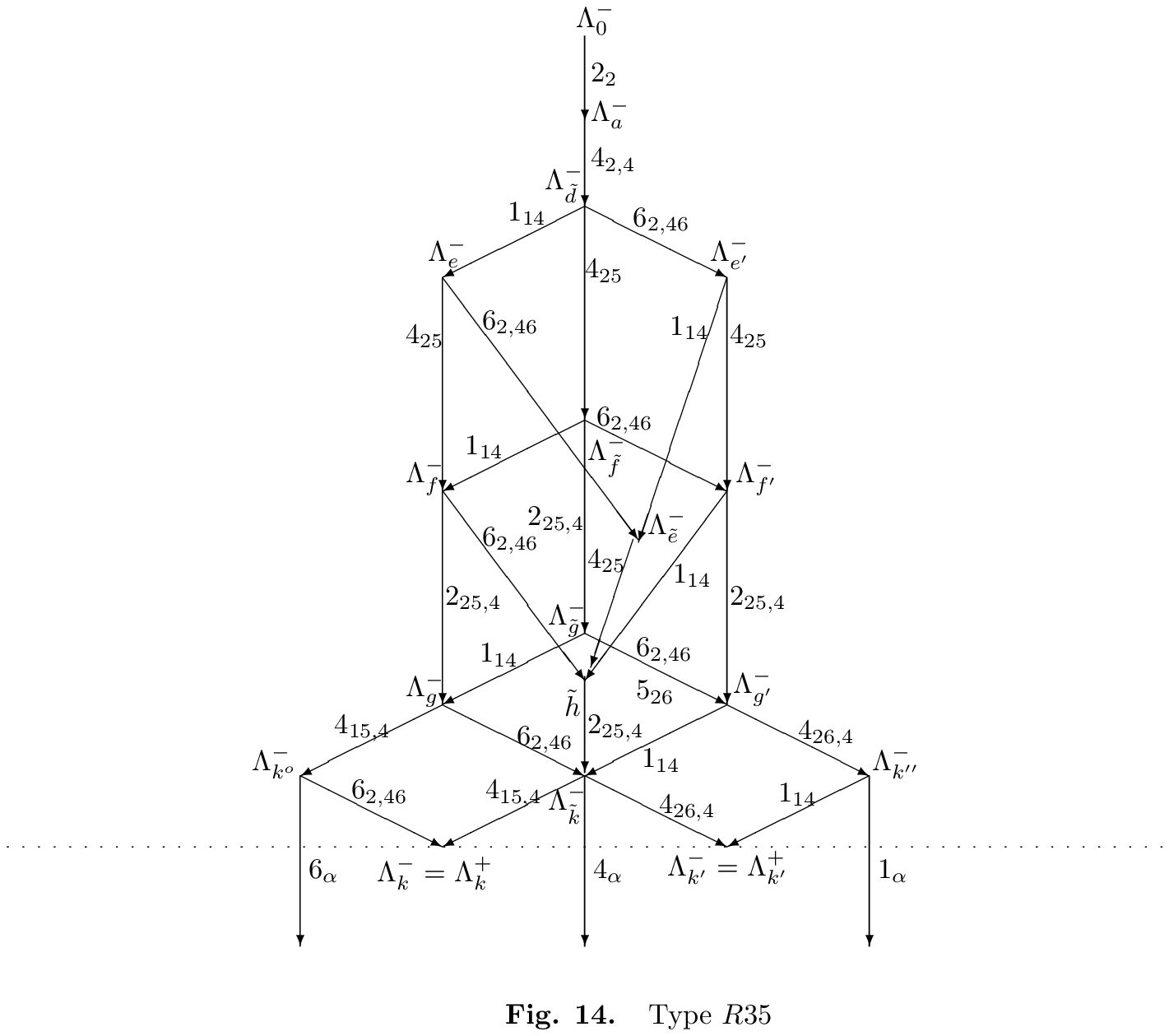}{3.8cm}
\fig{}{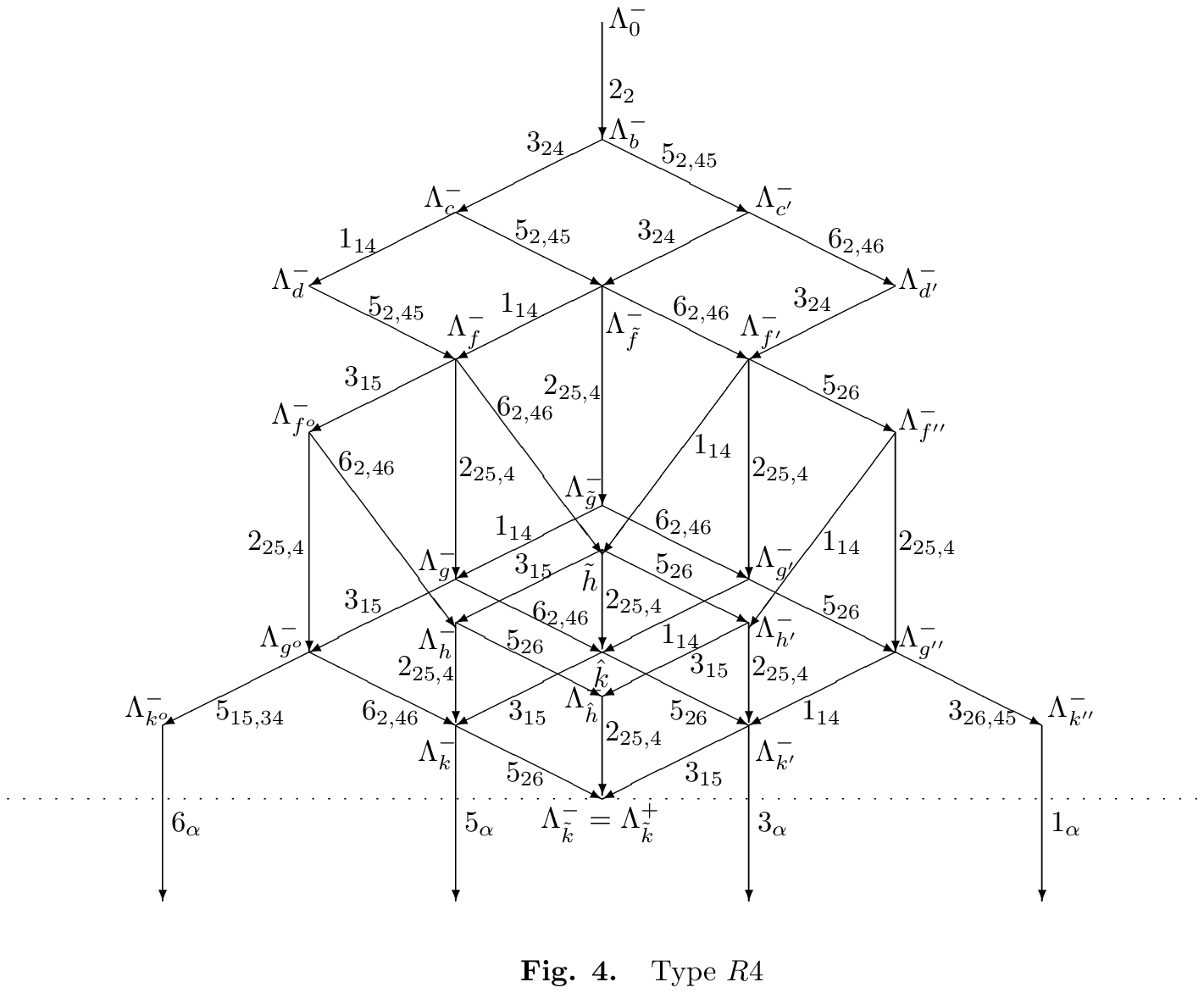}{5.4cm}
\fig{}{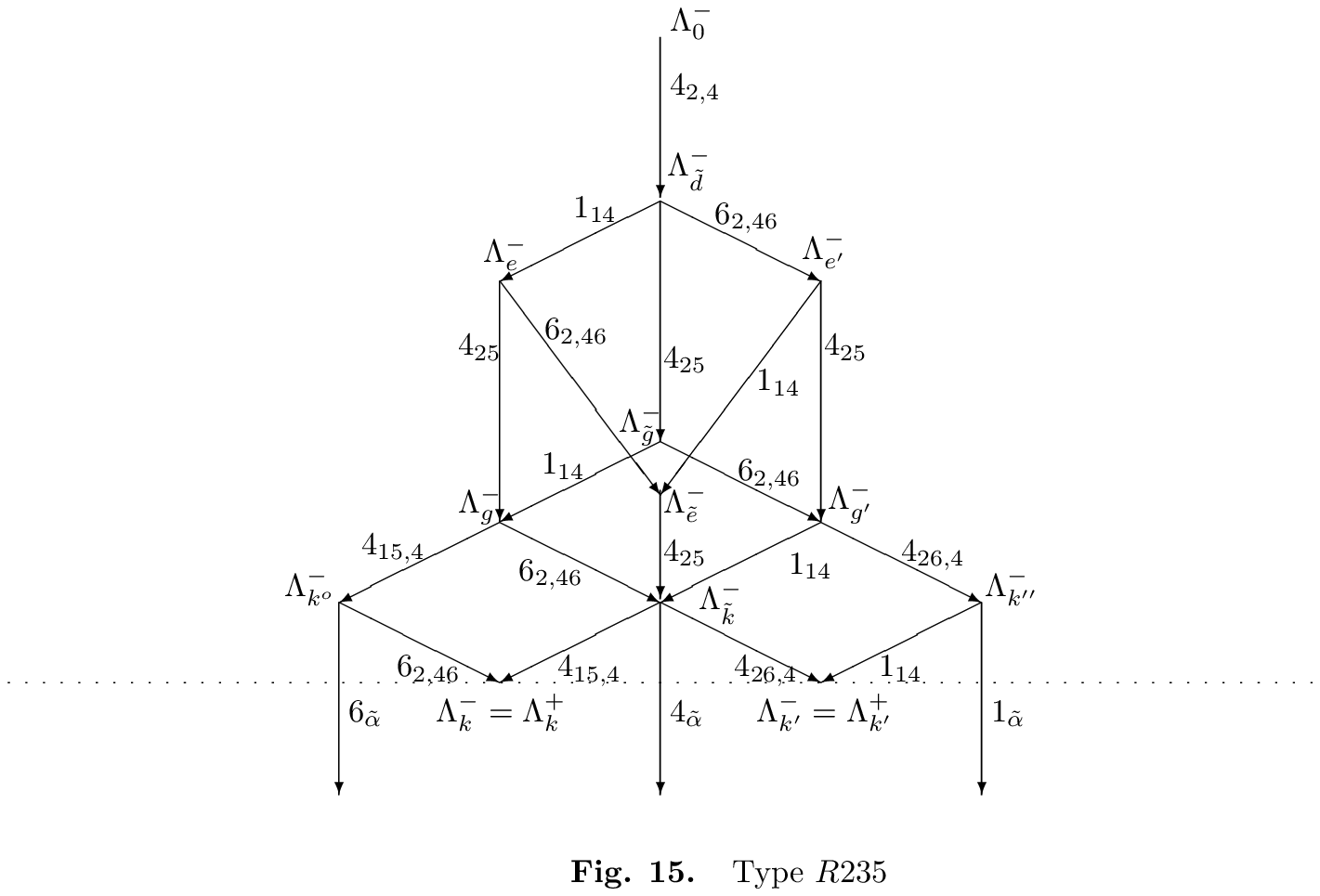}{3.8cm}

\np

\fig{}{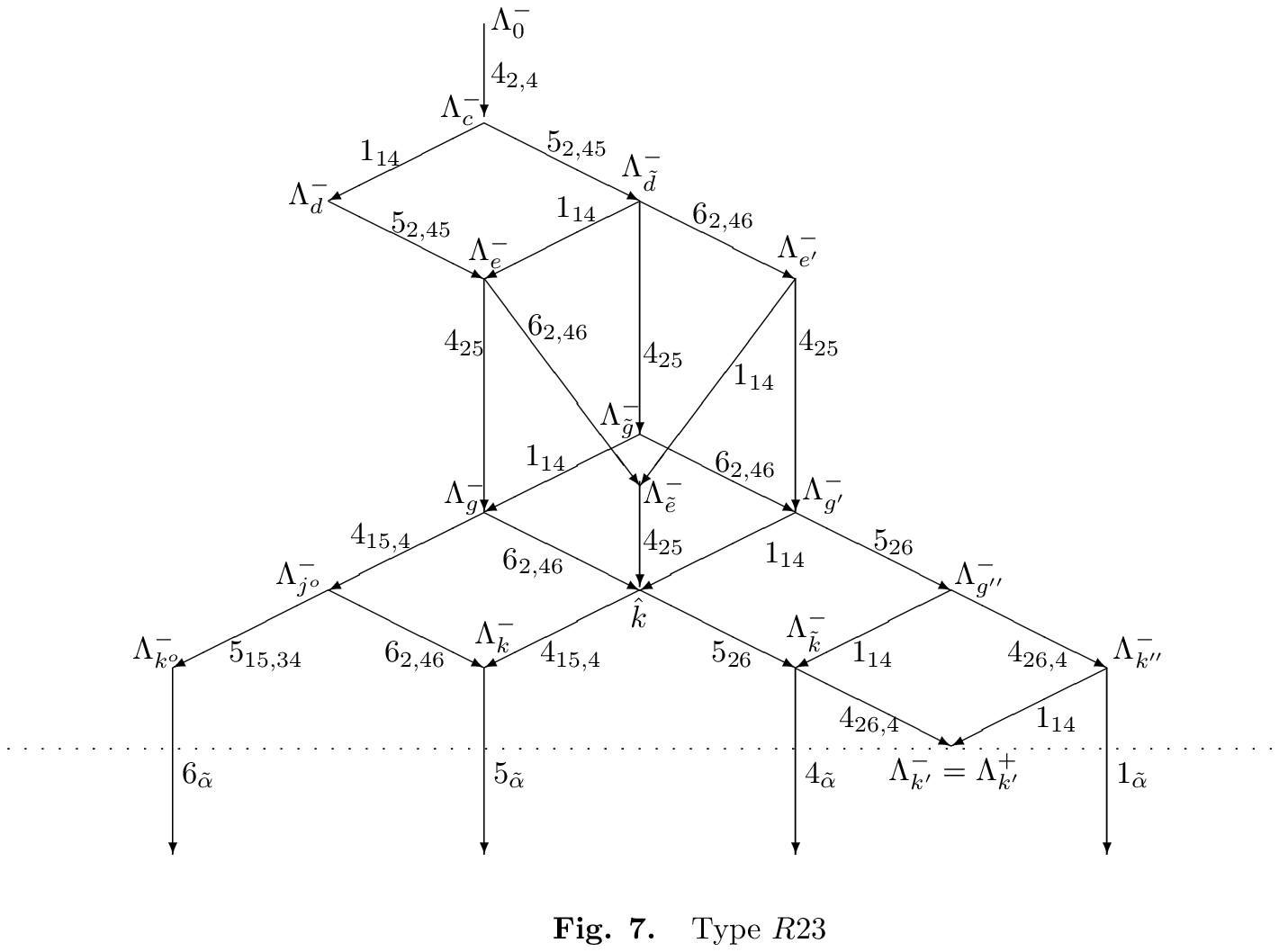}{5.4cm}
\fig{}{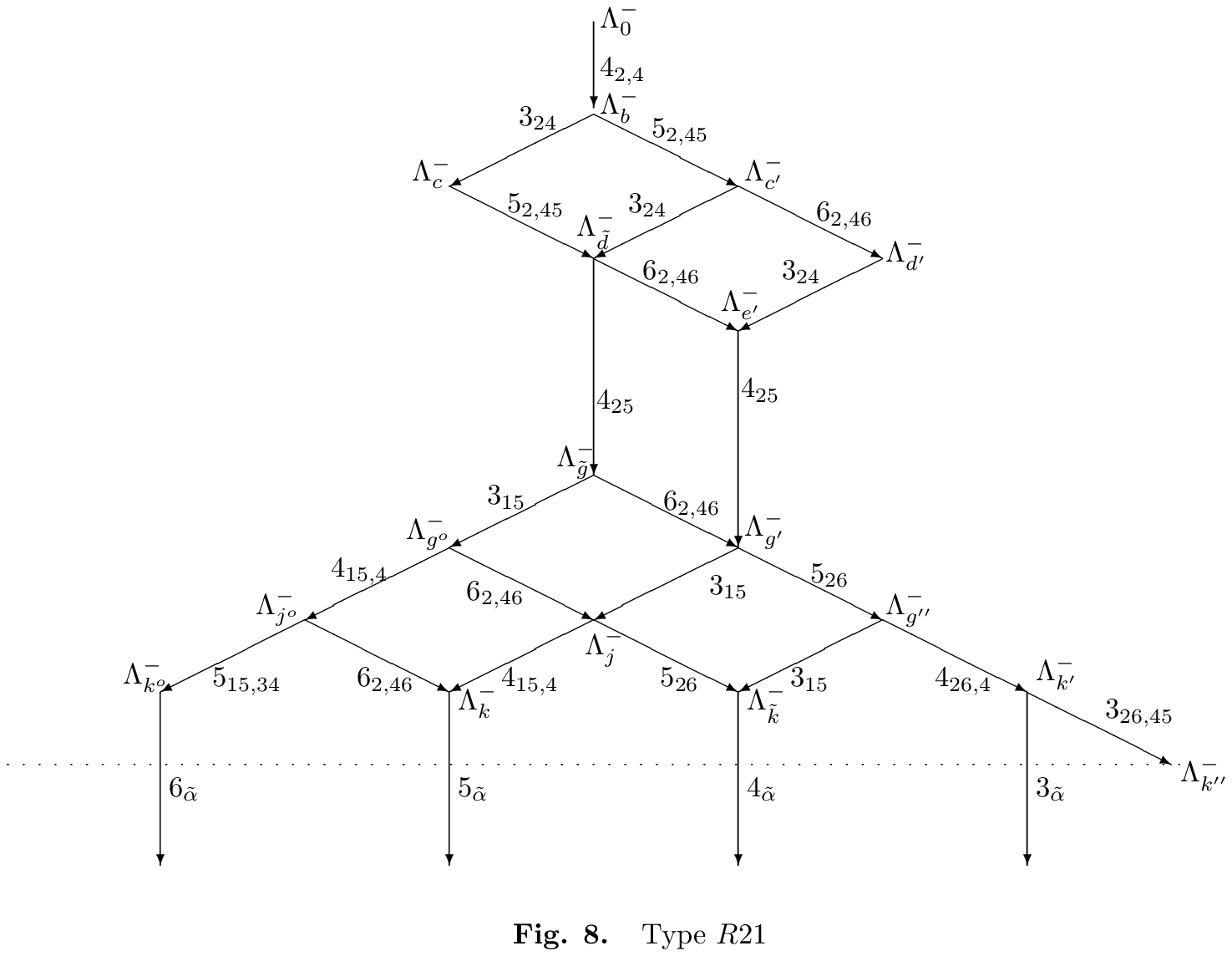}{6.1cm}
\fig{}{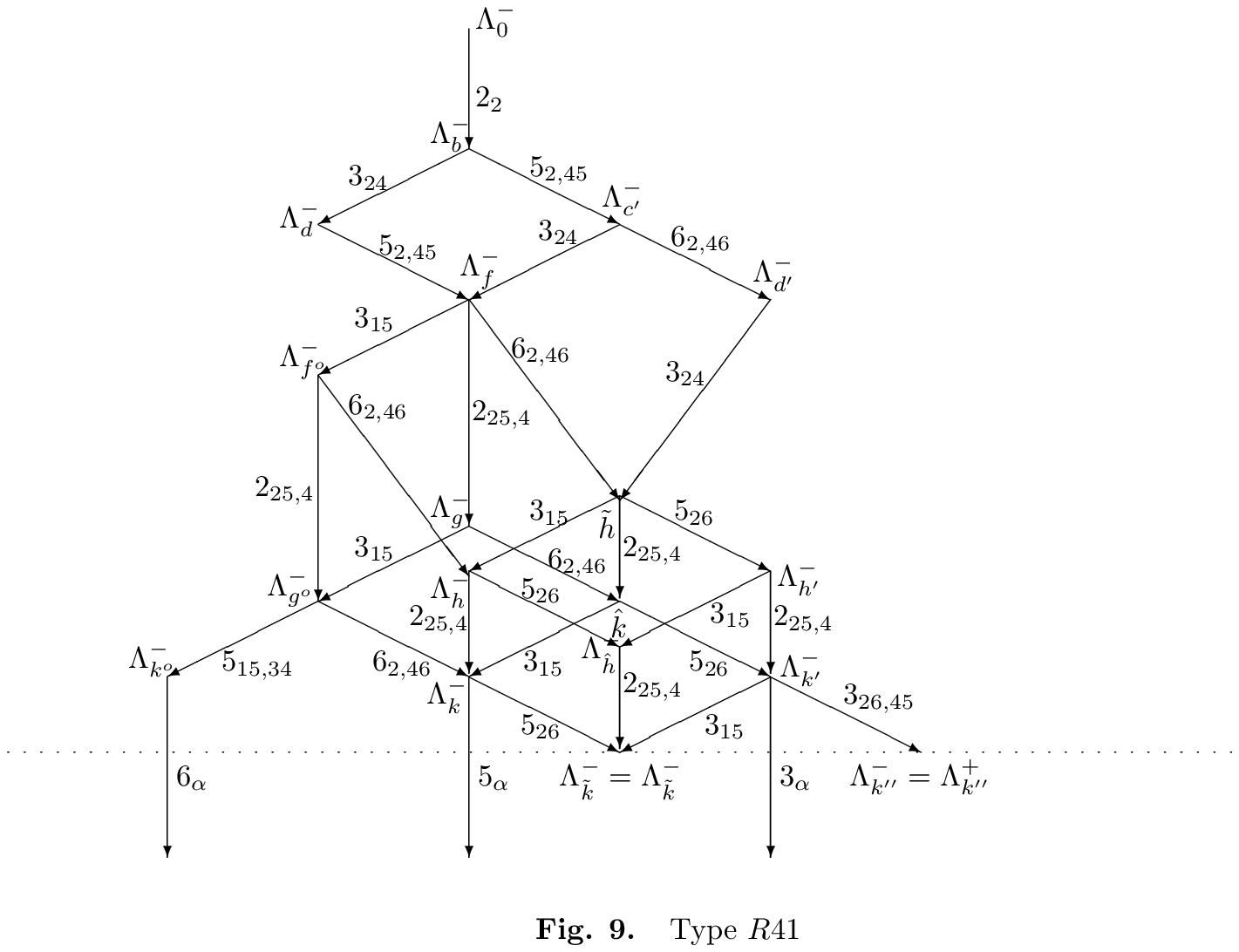}{4.7cm}

\np

\fig{}{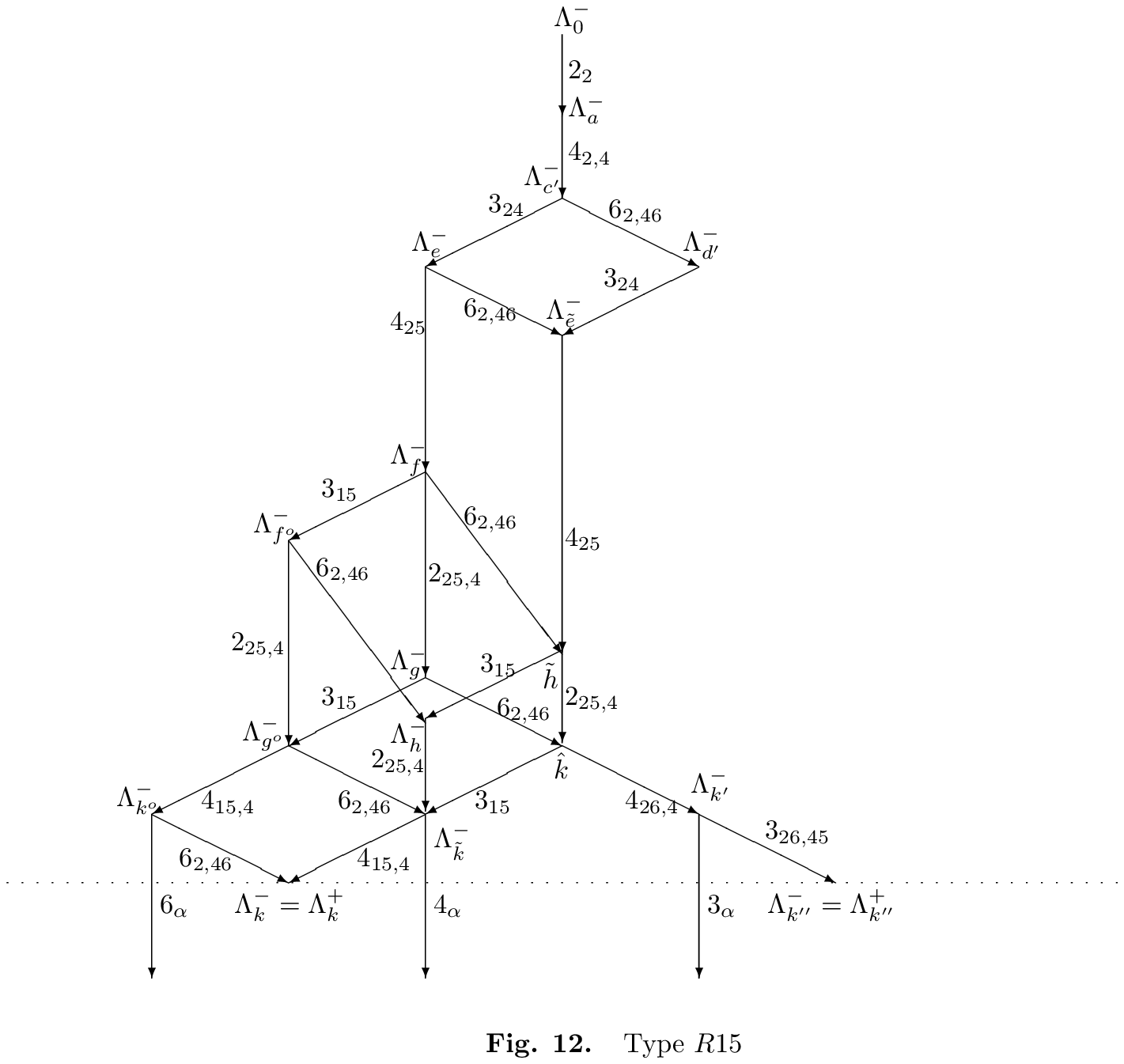}{4.7cm}
\fig{}{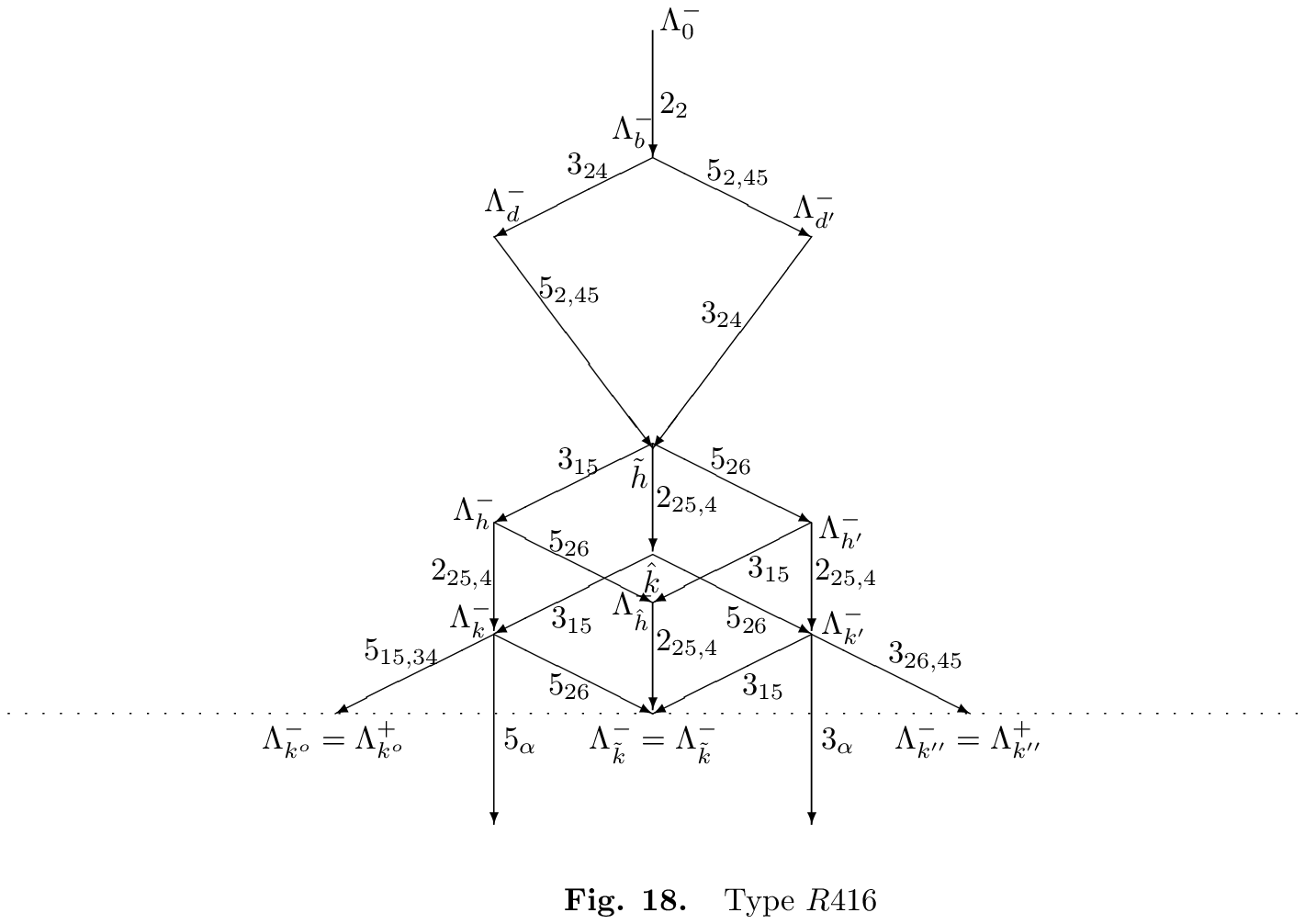}{4cm}
\fig{}{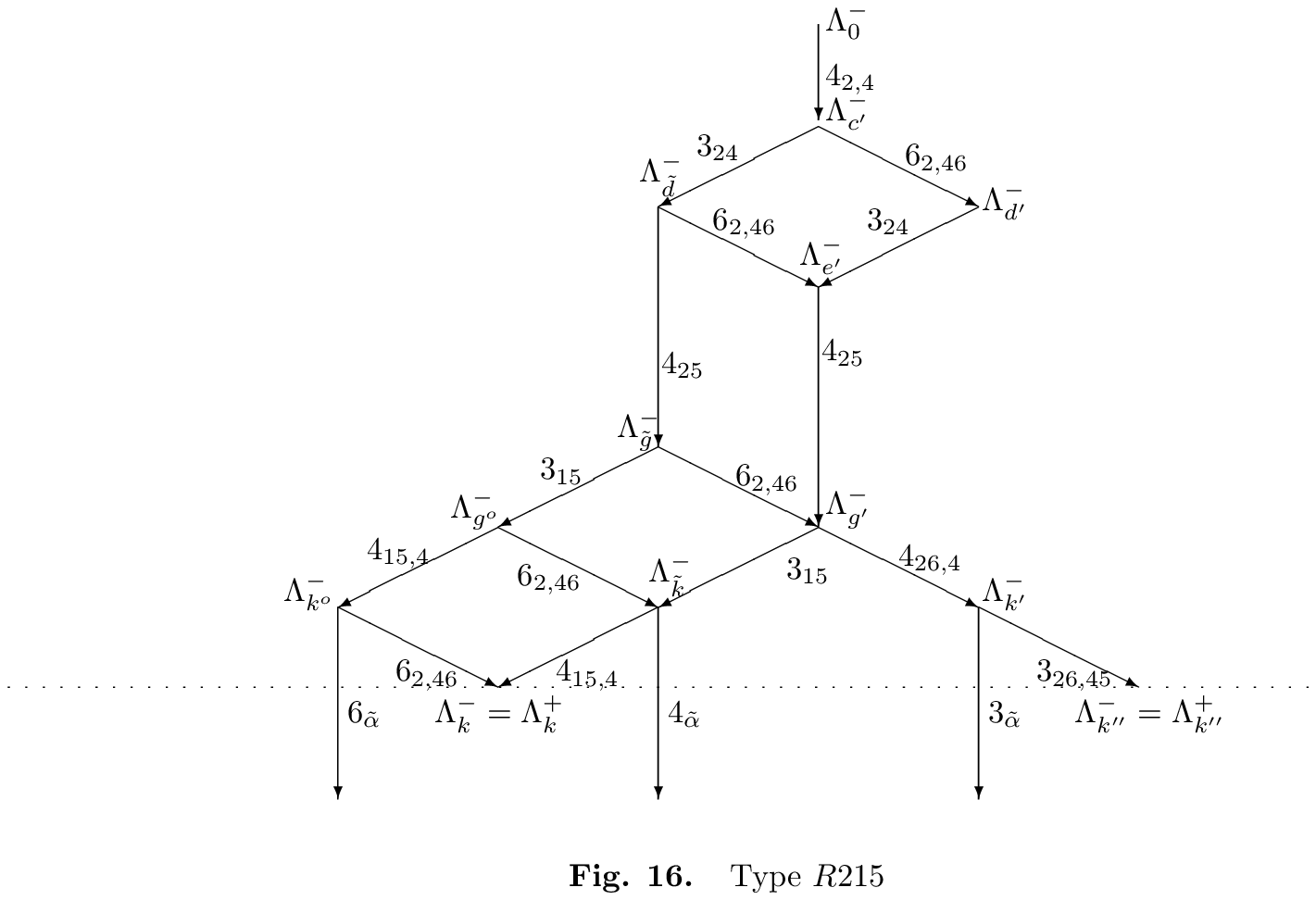}{4.7cm}

\np

\fig{}{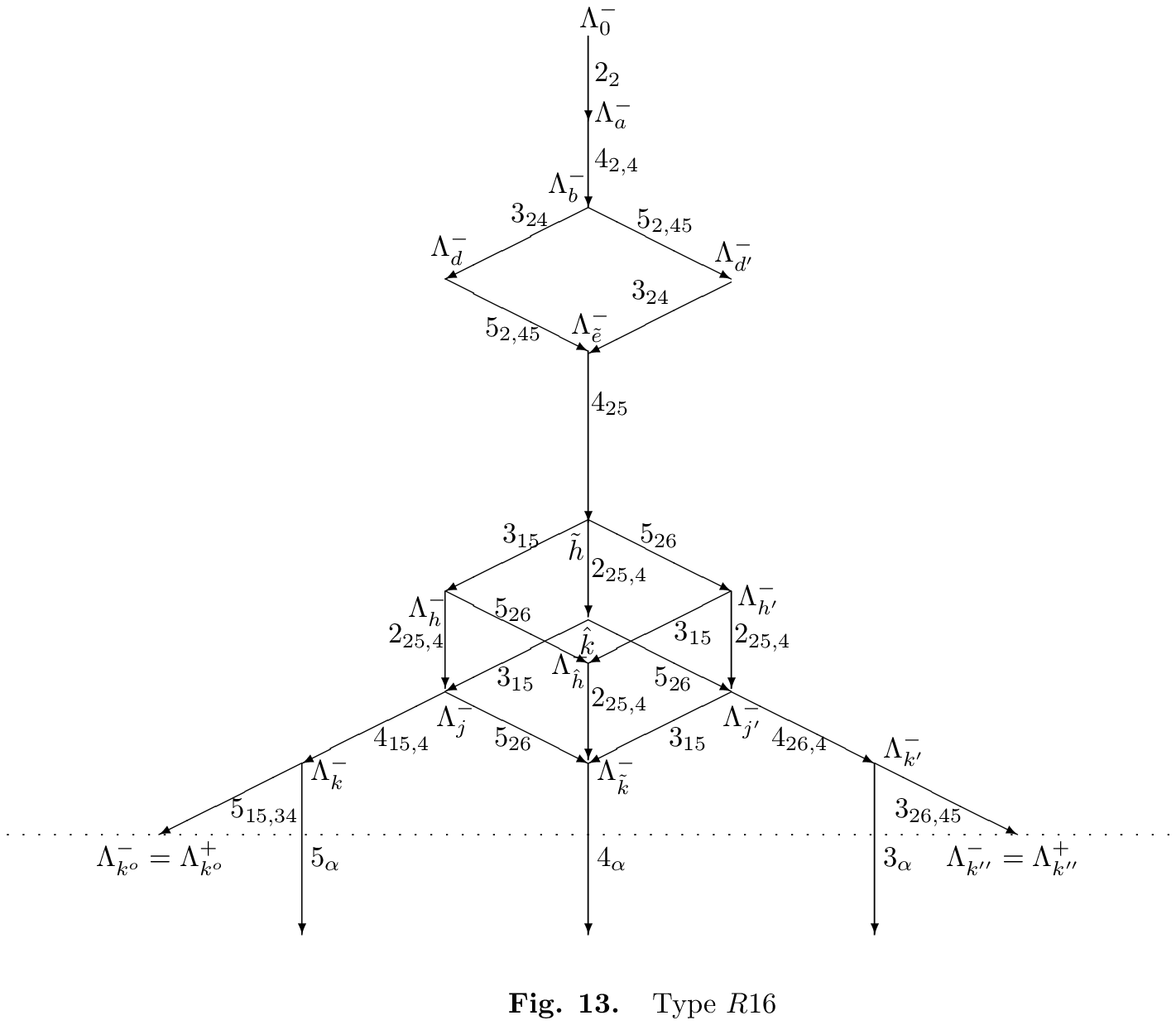}{5.4cm}
\fig{}{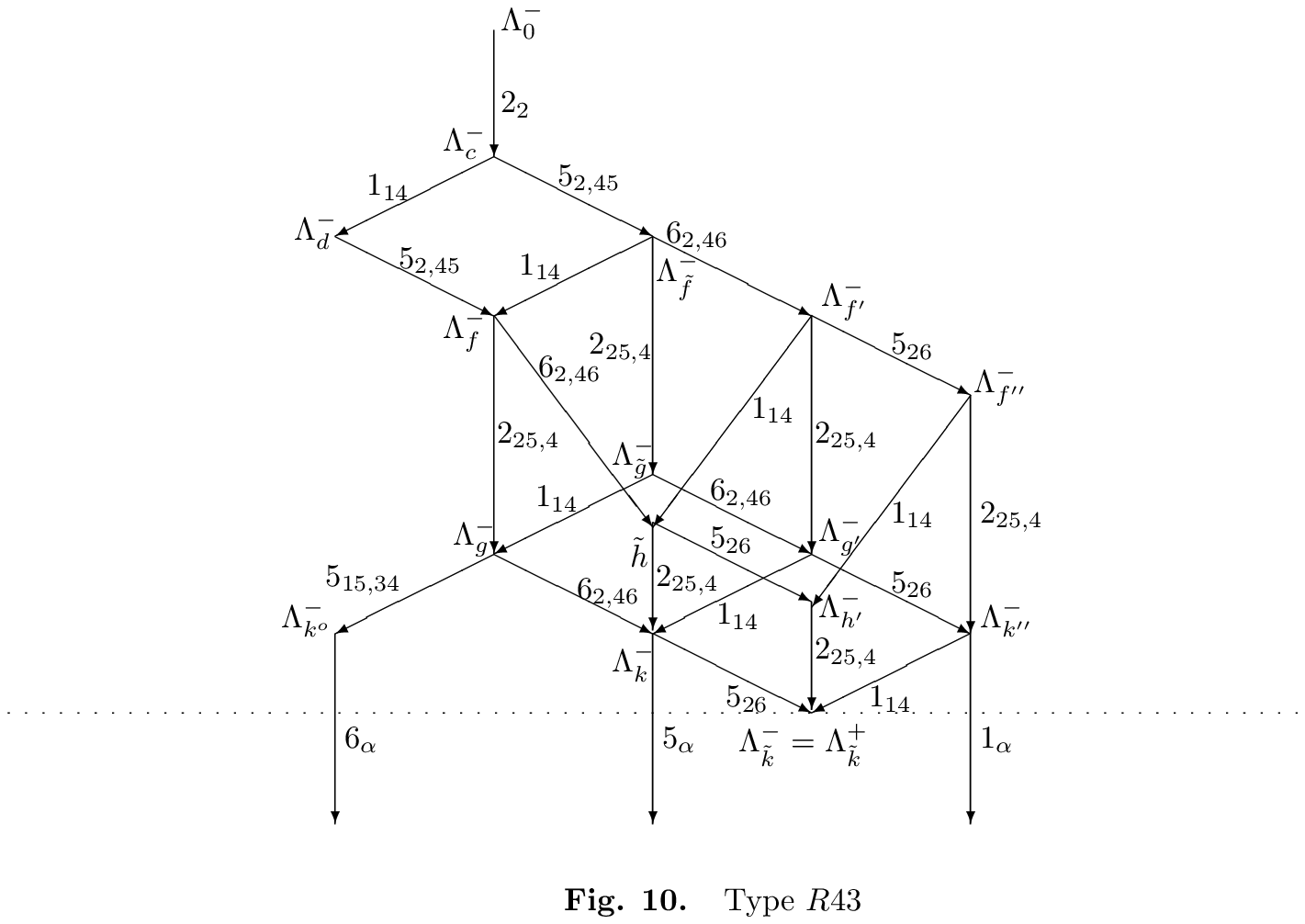}{4cm}
\fig{}{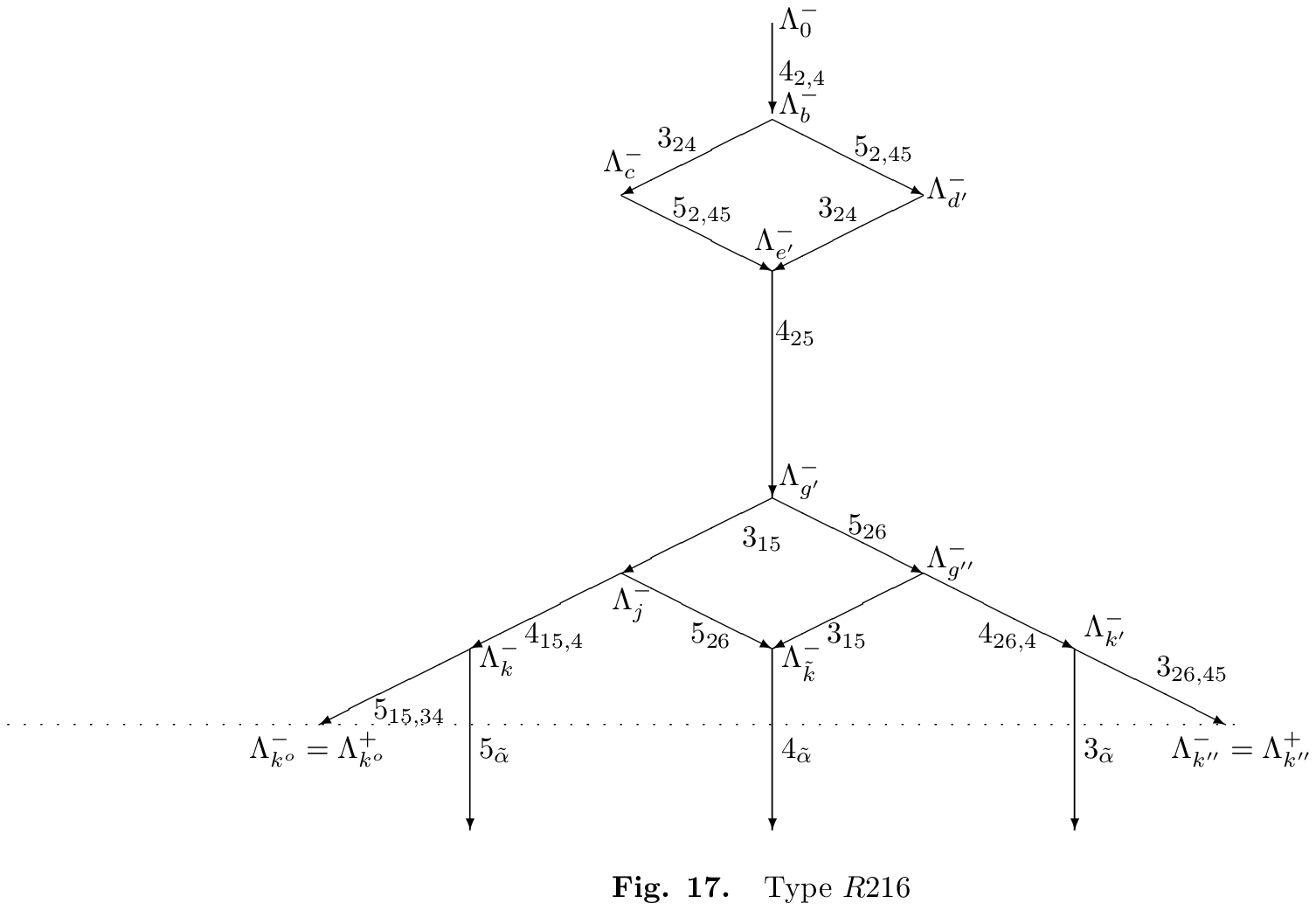}{5.6cm}

\np
\end